\documentclass[acmsmall]{acmart}

\usepackage{subcaption}
\usepackage{caption}
\usepackage{xcolor}
\usepackage{longtable}
\AtBeginDocument{%
  \providecommand\BibTeX{{%
    \normalfont B\kern-0.5em{\scshape i\kern-0.25em b}\kern-0.8em\TeX}}}

\setcopyright{acmcopyright}
\copyrightyear{2018}
\acmYear{2018}
\acmDOI{10.1145/1122445.1122456}

\acmJournal{JACM}
\acmVolume{37}
\acmNumber{4}
\acmArticle{111}
\acmMonth{8}

\newcommand{\hl}[1]{{\color{black}#1}}
\begin{document}

\title{A Method to Analyze Multiple Social Identities in Twitter Bios}

\author{Arjunil Pathak}
\affiliation{%
  \institution{University at Buffalo}
  \city{Buffalo, NY}
  \country{USA}}
\email{arjunil.pathak@gmail.com}

\author{Navid Madani}
\affiliation{%
  \institution{University at Buffalo}
  \city{Buffalo, NY}
  \country{USA}}
\email{navidmdn74@gmail.com}

\author{Kenneth Joseph}
\affiliation{%
  \institution{University at Buffalo}
  \city{Buffalo, NY}
  \country{USA}}
\email{kjoseph@buffalo.edu}
\renewcommand{\shortauthors}{Anonymous et al.}

\begin{abstract}

Twitter users signal social identity in their profile descriptions, or bios, in a number of important but complex ways that \hl{are not well-captured by} existing characterizations of how identity is expressed in language. Better ways of defining and measuring these expressions may therefore be useful both in understanding how social identity is expressed in text, and how the self is presented on Twitter. To this end, the present work makes three contributions. First, using qualitative methods, we \hl{identify and} define the concept of a personal identifier, which is more \hl{representative} of the ways in which identity is signaled in Twitter bios. Second, we propose a method to extract all personal identifiers expressed in a given bio. Finally, we present a series of validation analyses that explore the strengths and limitations of our proposed method. Our work opens up exciting new opportunities at the intersection between the social psychological study of social identity and the study of how we compose the self through markers of identity on Twitter and in social media more generally.
\end{abstract}



\maketitle

\section{Introduction}


A \textit{social identity} is a word or phrase that refers to a particular social group (e.g. ``Yankees fan'' or ``athlete''), category (``woman'') or role (``parent'') \cite{smith-lovin_strength_2007-1}. Social identities are important because people select identities for themselves and others in social situations and then use those identities to determine how to behave \cite{ballDefinitionSituationTheoretical1972}. We select different identities for ourselves and others at work versus at home, for example, and behave differently in these settings because of it. A vast literature in social psychology explores various properties of social identities, considering, for example, their affective meanings and links to behavior \cite{heise_affect_1987}. Much is therefore known about, for example, how people feel about professors, who society thinks should be a professor, and how professors are likely to act.

Social media presents a potentially rich resource to refine and expand our understanding of social identity \cite{marwick_i_2011-1,devito_how_2018-1}.  For example, on social media, our many social circles are entangled in a single audience, presenting a unique opportunity to study how people navigate situations in which they may want to take on multiple identities at the same time \cite{marwick_i_2011-1}.  Social media also present new ways of measuring behaviors and structures associated with different identities. We can observe\hl{, for example,} what kinds of social networks users who present particular identities are embedded within, and the language they express \cite{bamman_gender_2014}. Existing literature on identity can also help us better understand \hl{online behavior}. For example, we can use meanings that researchers have attached to particular identities via survey analysis to predict how users who identify in certain ways are likely to behave on platforms like Github \cite{morgan2019modeling}.

Realizing these links between studies of social identity and social media involves, of course, the ability to determine the social identities being presented by users. Increasingly, scholars have turned to profile descriptions, or \emph{bios}, to do so. For example, bios have been used to find individuals \hl{who identify} as journalists \cite{zeng2019detecting}, to study the evolution of self-presentation through political identities \cite{rogers2021using},  and to study how users navigate contested social categories like gender \cite{haimson_constructing_2016}. Other\hl{s have} shown that bios also contain rich information that moves beyond social identity.  For example, personal interests do not fit neatly into a group, category, or role, but are ways that people often identify themselves in bios on tumblr \cite{yoder2020phans}. 

More broadly, social media bios provide a wealth of information about the identities that users have chosen to present, but do so via words and phrases (and @mentions and hashtags) that do not always neatly align with standard assumptions about how social identities are expressed in language.  Consider, for example the bios of \hl{current} U.S. President Joe Biden and \hl{current} U.S. Congresswoman Alexandria Ocasio-Cortez:
\begin{quote}
    46th President of the United States, husband to @FLOTUS, proud dad \& pop. Tweets may be archived: http://whitehouse.gov/privacy \\\\
    US Representative,NY-14 (BX \& Queens). In a modern, moral, \& wealthy society, no American should be too poor to live. 100\% People-Funded, no lobbyist \$. She/her.
\end{quote}
\hl{President} Biden's bio contains complex, lengthy identities like ``46th President of the United States'' and identities that are signaled relative to other people (husband to @FLOTUS). It also contains phrases that have no clear relationship to any identity (the phrase ``Tweets may be archived''). \hl{Congresswoman} Ocasio-Cortez's bio contains important markers of her identity, e.g. pronouns, a statement that she is ``100\% People-funded'', and a concise statement of her political ideology. These latter two phrases are  \emph{identifying}, in that they imply particular (set of) identities that Ocasio-Cortez associates with.  But they are not, by the commonly used definition stated at the beginning of this paper, social identities.  


The goal of the present work is to provide a better understanding of the\hl{se} rich \hl{and complex} ways in which individuals express social identity in social media bios. We also aim to provide evidence of the possibilities and challenges in developing \hl{automated} methods to do so.  In particular, the present work focuses specifically on Twitter bios, and makes three main contributions:
\begin{itemize}
    \item {\bf Using qualitative methods, we develop the concept of a \emph{personal identifier}, which better encapsulates the many ways users express identity in Twitter bios.}  We define a personal identifier as \emph{a word or phrase which, when read, can be used to make a direct assumption about a social identity (or set of social identities) held by an individual without any additional context, and where it is likely that the user intends to present that inferred identity}. We perform a series of annotation studies that show that personal identifiers can be identified with moderate reliability (Krippendorf's alpha of .6-.8 across various annotation tasks), signifying \hl{the concept's} potential to be used meaningfully for downstream research questions
    \item {\bf We develop a method to extract personal identifiers from Twitter bios} - Our method uses a single regular expression to split a bio into a series of delimited phrases, which are then cleaned and filtered to generate a list of personal identifiers within that bio.
    \item {\bf We evaluate the \hl{reliability, validity, utility, and generalizability} of using phrases extracted from Twitter bios to study social identity} - We provide a series of analyses to assess our method to extract personal identifiers from Twitter bios. Overall, we find that we can extract valid and useful measures of presented social identity in Twitter bios, but that questions remain about the generalizability of our approach to samples on which we did not perform qualitative analysis.
\end{itemize}

Our work is of interest to social psychologists who study social identity and its implications, and \hl{to} CSCW scholars who study the social media self. In the former case, we provide a novel data source for the study of which individuals present which identities, and which identities are often combined together to present a single self concept. In the latter case, we provide a new way to characterize \hl{how} the self is presented within the context collapse of social media. All code used in the present work can be found on Github, at \url{https://github.com/kennyjoseph/twitter_personal_identifiers}, as well as in the Supplemental Material for this submission.

\section{Background}

\subsection{Identity and Language}

Scholars have long studied the ways in which identity and/or the self are represented in language \cite{bucholtz_identity_2005,heise_affect_1987,johnstone_linguistic_1996}.  For example, narrative scholars use qualitative methods to analyze complex identities expressed in the text of personal narratives \cite{riessman2003analysis}, and social psychologists have proposed text-based methods to study the relationships between dictionary definitions of pre-defined lists of social identities \cite{heise_self_2010-1}. Our work aligns best with the latter line of research, in which we assume identity can be characterized via \hl{(}a set of\hl{)} phrases. However, our work is an effort to use modern methods and data to help link these two lines of work by moving beyond the standard, restrictive definition of what constitutes \hl{an} expression of a particular social identity in text. In doing so, we follow recent calls in the CSCW literature \cite{saxena2020methods,fiesler2019qualitative,antoniak2019narrative} and elsewhere \cite{nelson_computational_2020} to bridge qualitative and quantitative methods. 
 
Quantitative social psychological work on social identity makes two often unstated, but nonetheless critical, assumptions. First, it is generally assumed that we know, a prior, the possible words and phrases that people use to describe themselves and others \cite{joseph_exploring_2016}. Second, we assume that these words and phrases align with the definition of social identity given above. For example, the growing literature in computational social science focusing on measuring stereotypes using natural language processing focuses almost exclusively on small sets of pre-defined lists of social identities  \cite{ahothali_good_2015,kozlowski_geometry_2018,garg_word_2018,zhao_gender_2018,bordia_identifying_2019,may_measuring_2019}.  The present work challenges both of these assumptions - that we should be focused only on explicit references to social identities, or that even with this focus, pre-defined lists can be constructed that capture all relevant identities.

Our work therefore ties into literature that explores other ways in which the self is presented in language. One line of this work focuses on how people holding different identities use language differently \cite{cameron1998performing,becker_linguistic_2014}. For example, Bamman et al. \cite{bamman_gender_2014} study how gender identity is expressed on Twitter via language in tweets.  
More directly related to the proposed work is research studying how people talk \emph{about themselves}, and in particular, how they do so in social media bios \cite{shima_when_2017,rogers2021using}. Scholars have used bios on location-based dating apps to explore how expressions of stigmatized and localized identities emerge within them \cite{birnholtz_identity_2014-1}.  Others have assessed how affective presentations of the self emerge as individuals navigate gender transitions \cite{haimson_constructing_2016}. \citet{priante_whoami_2016} use a theoretically informed approach to understand the types of selves presented, differentiating between, for example, relational and occupational selves.  \citet{semertzidis_how_2013-1} provide an overview of the language present in Twitter profile descriptions, finding evidence that users identify themselves with occupations, interests, and ``personal info.''

Most similar to our work are three recent efforts.\citet{rogers2021using}, motivates the use of Twitter bios specifically as a means of presenting one's identity, and shows how a hand-selected set of political identities evolved over time in Twitter bios. \citet{yoder2020phans} use regular expressions to extract out various ways in which individuals express their identity in profile descriptions on tumblr. They then use these to better understand patterns of homophily and heterophily in reblogging.  Finally, \citet{li-etal-2020-emoji} focus specifically on emojis in bios, showing patterns in co-occurrence of emojis how the emojis used predict various components of the tweets users send, and that users are more likely to follow others with the same emojis.

Our work compliments this growing literature in two ways. First, we provide a general concept, the personal identifier, that encompasses many of the ways people express social identity within bios. Second, we draw connections between personal identifiers extracted from bios and quantitative social psychological work that measures properties of specific identities, showing how these two can be usefully combined.

\subsection{Self-Presentation and Social Identity}

Scholars have developed many ways to help understand the self as it exists in online contexts (e.g.  
\cite{lampe_familiar_2007,lim_online_2015,thorne_technologies_2015,marder_strength_2016,ranzini_love_2017,tandoc_platform-swinging_2019,marwick_i_2011-1}). Central to this vast literature is a relatively simple question--- what determines, or constrains, the (set of) identities that an individual will take on in a given context?  

One set of constraints revolves around limits on which identities are available to us. For example, institutionalized and/or culturally normative structures of inequality can prevent women and Black individuals from obtaining the same high status occupational identities as men \cite{blau_presidential_1974}. These constraints also emerge denotatively in language - for example, the identity father is denotatively aligned with men.  In the present work, we use these well-established denotative and structural constraints as a form of convergent validity; that is, as a way to validate that the personal identifiers extracted from bios align with these constraints in expected ways.

Other constraints are much more dynamic and contextual, in that we take on different identities depending on where we are \cite{smith-lovin_self_2003} and who we are interacting with \cite{davis2014context}. 
 Social media blurs these boundaries, e.g. between public and private, professional and personal, and the many different selves and situations individuals find themselves in \cite{hanusch_journalistic_2017,molyneux_how_2017,marwick2011see}. This collapse of social contexts is what is referred to as \textit{context collapse} \cite{boyd2002faceted,marwick_i_2011-1}. Because of the context collapse, users often find the need to define themselves along \emph{multiple}, ``authentic'' dimensions \cite{marwick_i_2011-1}.
These decisions are influenced by not only the affordances of the platform, but how individuals \emph{understand} those affordances \cite{devito_how_2018-1,devito_platforms_2017}. Thus, individuals may make decisions that are consistent with their worldview, but potentially inconsistent with reality.

The present work presents a new tool to help understand how social media users, and Twitter users in particular, navigate both these structural and contextual constraints when presenting themselves in bios. We provide evidence that this tool is useful in two ways. First, we look at patterns in how individuals adopt multiple identities (e.g. professor and athlete) and what this can tell us about the ways that users navigate the context collapse. Second, we look at differences in affective, gendered, and personality-based perceptions of the identities that users do or do not adopt.

\section{Data}

\begin{table}[t]
    \centering
    \begin{tabular}{|p{6cm} p{2cm} p{2cm} p{2cm}|}  \hline 
     {\bf Dataset}    &  {\bf Panel 1} &  {\bf Panel 2} &  {\bf Random} \\ \hline 
     Collection Date & July 7th, 2020 & Sept. 4th, 2019 & Oct. 3rd, 2019 \\ \hline
     \multicolumn{4}{|c|}{{\bf Information on Filters Applied}} \\ \hline 
      \% Blank Bio  & 27.4\% & 25.3\% &  46.7\% \\ 
      \% Remaining Containing Org. Language  & 0.1\% & 0.1\% &  0.1\% \\
      \% Remaining Last Status Not English/Spanish/Unknown & 1.7\% & 1.16\% &  33.9\% \\
      \% Remaining Of Original Dataset in Final & 71.3\% & 73.4\% & 35.1\% \\ \hline
      Total N Bios Used for Validation & 1,173,834 & 1,141,493 & 526,188 \\ \hline
      \multicolumn{4}{|c|}{{\bf Statistics on Users \& Bios}} \\ \hline
      \% Verified & 0.8\% & 0.8\% & 0.2\% \\
      Median Followers & 115 & 127 & 75 \\ 
      Median Friends & 219 & 231 & 145 \\ 
      Median Status & 651 & 696 & 328 \\ 
      Median Profile Length (Characters) & 70 & 70 & 46 \\ \hline
      \multicolumn{4}{c}{{\bf Demographics of the Panel}} \\ \hline
      \% Identifying male (vs. female) & \multicolumn{2}{c}{55\% (9\% missing)} & - \\ \hline
      Average \% Rural & \multicolumn{2}{c}{15\% (5\% missing)} & - \\ \hline
      \% Black & \multicolumn{2}{c}{ 8\% (7\% missing)} & - \\ \hline
      \% Hispanic & \multicolumn{2}{c}{ 5\% (7\% missing)} & - \\ \hline
     \% Asian & \multicolumn{2}{c}{ 2\% (7\% missing)} & - \\ \hline
     \% White & \multicolumn{2}{c}{ 85\% (7\% missing)} & - \\ \hline 
     \% Registered Democrat & \multicolumn{2}{c}{ 64\% (64\% missing)} & - \\ \hline 
    \end{tabular}
    \caption{Summary information for the three datasets studied in the present work. The table provides the date on which the bios were collected, effects of the steps used to filter the bios, and statistics pertaining to users and bios after filtering was performed. Finally, we provide demographic information about the Panel 1 and Panel 2 datasets (we do not have this information for the Random dataset). For demographics, we provide the same information across samples for ease of presentation (these numbers were consistent for both samples), and include the percentage of users that were missing data.}
    \label{tab:data}
\end{table}

\subsection{Overview}

Table~\ref{tab:data} outlines basic statistics of the three datasets of Twitter bios used in this work. We used one primary dataset, which we call \emph{Panel 1}, to conduct the qualitative analyses described below and to develop our algorithm to extract personal identifiers.  This dataset contains 1,174,834 bios (after filtering, see below) from a \emph{Panel} of Twitter users that have been linked to voter registration records. See Section~\ref{sec:linking} below for details on this linking process. In order to explore the generalizability of the proposed approach, we also conduct a subset of our validation analyses on two additional datasets. The first, \emph{Panel 2}, begins with the same set of users as in \emph{Panel 1}, but with bios collected approximately 10 months prior. Almost 81\% of users had the exact same bio between these two periods, but we still find it useful to explore potential issues with generalizability over time. We also analyze bios of a set of users we refer to as the \emph{Random} sample, a collection of Twitter accounts sampled in a similar fashion to \emph{Panel 1} and \emph{Panel 2}, but that are not linked to voter registration records.

All three samples were constructed by first identifying a set of approximately 400M Twitter users who sent at least one tweet in the Twitter decahose some time \hl{between} 2014 \hl{and} 2016. The bios of these users were then re-collected at a later time (see Collection Date in  Table~\ref{tab:data}).  In all samples, accounts were then filtered, in order, by 1) removing accounts with blank bios, 2) removing bios with obvious indications of being linked to an organization (containing the terms ``we are'' or ``not affiliated'' \cite{wood-doughty_predicting_2018}), and 3) removing accounts where \hl{we could identify a} last sent tweet by the user and 2) for a language other than English or Spanish, according to the Twitter API. The percentage of profiles removed for each of these filtering steps is shown in Table~\ref{tab:data} for each dataset.

In addition to information on these filtering steps, Table~\ref{tab:data} provides summary statistics on the users and bios in the final datasets used for analysis here.  For the datasets linked to voter registration records, demographic information is also provided (see below for how this was obtained). Readily apparent from \hl{Table~\ref{tab:data}} are the expected similarities between the two different collections of the same set of Twitter users (Panel 1 and Panel 2), and the vast differences between the Panel datasets and the Random dataset. Users in the final Panel datasets are 30 times less likely to have a non-English, non-Spanish last tweet \hl{(before filtering)}, 4 times more likely to be verified, and have sent nearly twice as many tweets (statuses). These differences will be important in understanding differences in results across the Panel and Random datasets.

\subsection{Linking to Voter Records}\label{sec:linking}

An interest in self-presentation necessitates a focus on people, and not, for example, bots or organizational accounts. Because of this, we would like to study only Twitter accounts we can be reasonably certain are personal Twitter accounts. In addition, as discussed below, one useful form of validation we seek is to confirm that identities relate to user demographics in expected ways. To address both of these data needs, the present work focuses in large part on Twitter accounts that have been linked to voter registration records.

The methods we use to link Twitter accounts to voter records are the same as those rigorously outlined and validated in work by \citet{grinberg_fake_2019} and \citet{hughes2020using}, and the same set of users as has been used in work since \cite{shugars2021pandemics}. Given our reliance on prior work, we provide only an outline of the methodology here, and refer the reader to their work for full details. We begin with a set of approximately 300 million voter registration records from TargetSmart,\footnote{The data was provided under a strict data-sharing agreement. No data that identifies individuals within the data can be shared with individuals not directly tied to this data-sharing agreement} and the set of 400M users above who shared at least one tweet that appeared in the decahose from 2014-2016. We then link a Twitter account to a voter registration record if a) the first and last names of the two records match, b) the location (city and/or state) of the two records match, and c) there is no other Twitter user that both has this first and last name and that has no stated location.  Our algorithm matches approximately 0.5\% of the voter registration records from TargetSmart. 

Based on rough estimates of the number of average daily Twitter users\footnote{\url{https://www.oberlo.com/blog/twitter-statistics}}, we estimate this dataset to contain somewhere between 1-3\% of all voting-eligible Americans on Twitter. This name-based linking methodology has been shown to produce panels of Twitter users that are similar, demographically, to the general U.S. population that uses Twitter \cite{hughes2020using}, and has high accuracy (over 90\%) in terms of the matched individuals appearing to be linked to the correct registered voter \cite{grinberg_fake_2019}.

The voter records we link Twitter users to contain information on their registered political party, sex, age, race, and city and county of residence. Notably, because many states do not require individuals to state their race on voter records, our measure of race is partially self-reported, and partially imputed. Finally, we also use the county of residence field to link users to census data, which we in turn use to provide information on the percentage of people within the user's county of residence that live in a rural area. Additionally, we have a large amount of missing data for party registration, due to the fact that most states do not collect this information from voters, and TargetSmart does not impute it.

\subsection{Ethical Considerations}
 
Our approach to linking Twitter users and voter records has been approved by the IRB at \hl{Northeastern} University. We do not seek to infer user's names or locations in any way, and as such anyone who does not specify these details fully in their voter record or Twitter profile is not included in our dataset. For this reason, our approach also does not violate Twitter’s Terms of Service.  

Still, there are at least two ethical considerations that are worth identifying here. First, the Panel dataset we use identifies individuals along problematic demographic categorizations, e.g. binary sex, and uses imputation of others. There is ample literature on problems with the former categorization structure (see \citet{spiel2019better} for a CSCW-oriented discussion). We provide a further discussion of this point below. Second, linking these two datasets poses a potential risk of re-identification. In the present work, we have minimized this danger by only presenting single phrases within bios, and also by focusing only on phrases that are common on Twitter.

\section{Methods}

Our methodology has three components, each described in separate subsections:
\begin{enumerate}
    \item Qualitative coding, resulting in the notion of a personal identifier and measures of reliability of that concept
    \item  A method to extract potential personal identifiers from bios
    \item  \hl{Strategies} to assess the validity, utility, and generalizability of our method 
\end{enumerate}
Encouraged by other recent work in text analysis \cite{nelson_computational_2020,muller2016machine,eads2020separating}, we opt to iterate through the processes of algorithm development, qualitative analyses, and validation. We focus here primarily on the results of that process, that is, the methodology resulting from the final iteration of this development cycle.


\subsection{Coding for Expressions of Identity - Personal Identifiers}\label{sec:pi_extract}

The three authors of the paper conducted several rounds of qualitative coding, first on complete Twitter bios, and then, as we developed our \hl{approach}, on phrases extracted from Twitter bios. Our qualitative coding was conducted with a focus on understanding the ways in which Twitter users express social identity within bios, and how these expressions could be organized into a single, reliable concept.

Our earliest rounds of analysis were geared towards a better understanding of the data, and as such we did not compute reliability statistics. We instead focused on discussion and interpretation collaboratively amongst authors \cite{mcdonald2019reliability}. In these rounds, we quickly settled on the idea, first put forth by \citet{yoder2020phans}, that Twitter users often place \emph{explicit delimiters}, like commas, periods, and vertical bars, to separate out different ways of identifying themselves. This idea became central to our algorithm to extract personal identifiers. We also made a decision early in our process to avoid consideration of emojis for this work. While emojis are a meaningful markers of identity \cite{li-etal-2020-emoji}, it was often hard to determine what constituted a single expression of identity when multiple emojis were listed together. Given our focus on detecting explicit, separate identity terms, we opted to leave investigation of this phenomenon to future work. Beyond emojis, we found that individuals often expressed things in their bios that were definitively social identities (e.g. dancer, athlete), or definitively not relevant to any social identity (e.g. ``views are my own''). However, a vast grey area \hl{also} existed between these two cases, where bios seemed to signify identity but not in ways that fit existing definitions of identity in language. 
Specifically, we observed four often overlapping ways in which identity was indirectly expressed:
\begin{itemize}
    \item {\it Preferences}- Many bios stated lists of preferences, e.g. ``I love basketball and soccer''. We found this to be not meaningfully different from identities like ``basketball lover,'' which were also common. 
    \item {\it Personal Descriptors}- People often used series of adjectives to describe themselves, as in ``I am kind, smart, and old.''  We found this to be not meaningfully different from identities like ``old person.''
    \item {\it Affiliations}- People often revealed identity through indirect references to affiliations with social groups or movements. For example, a common term in profiles in this study was ``\#maga'', which reveals an underlying identity by way of an association with a political movement\hl{.}
    \item {\it Actions}- People often expressed their interests not as noun phrases but as verbs, as in ``hooping'' or ``running''. Many of these revealed underlying identities, although others were so vague as to reveal little about the individual (e.g. ``hugging'').
\end{itemize}
 
Authors agreed that these four classes of phrases did not meet the formal definition of a social identity. That is, they are not \hl{words} or phrases 1) used to refer to a particular social group, social category, or social role that 2) hold a particular set of culturally-constructed meaning(s) \cite{heise_affect_1987,mccall_identities_1978,smith-lovin_self_2003}. However, authors also agreed that these words and phrases \emph{did} constitute important information being conveyed about the self within a bio. More specifically, they \emph{implicitly signal association with an identity}, without explicitly naming that identity.  We therefore considered these expressions to be important ways of signaling identity and of interest for analysis.

After discussion, we agreed to merge these various ways of expressing identity into a single concept of \emph{personal identifiers}. Our initial definition of a personal identifier was a word or phrase which, when read, \hl{could} be used to make a direct assumption about a social identity (or set of social identities) held by an individual. 
However, later rounds of coding suggested two necessary amendments to this definition. First, many phrases which were identifiers \emph{in context} were not easily recognized as such when extracted out from a bio by our algorithm. An example of such a phrase is ``formerly,'' which requires additional information (formerly of what?) to hold meaning about social identity. These kinds of phrases were incorrectly extracted by our algorithm due to the imperfect nature of the extraction process. We therefore amended our definition of personal identifier to require that the phrase have meaning on its own, without additional context.
Second, certain phrases had clear implications for social identity, but it was not easy to discern \emph{which} identity was being signaled. An example of such a phrase is ``god.'' The phrase clearly signals an association with some kind of religious identity, but it is not clear \emph{which} religious identity. We therefore amended our definition of personal identifier to require that the phrase not only have meaning out of context, but that the annotator could be reasonably certain \emph{of a particular identity that the user was seeking to present}. 

The final definition of personal identifier we selected was therefore {\bf  a word or phrase which, when read, can be used to make a direct assumption about a social identity (or set of social identities) held by an individual without any additional context, and where it is likely that the user intends to present that inferred identity.} This definition is clearly subject to interpretation. To address the extent to which it could be reliably annotated, we therefore completed two final rounds of coding in which we assessed inter-rater reliability. These coding rounds are described in detail below, because they were also used to validate that the phrases extracted from bios were actually personal identifiers. We provide reliability statistics in Section~\ref{sec:reliability}, separate from validation results, to emphasize that reliability is a part of the research process \cite{mcdonald2019reliability}.

\subsection{A Method to Extract Personal Identifiers from Bios}

Our algorithm for extracting personal identifiers from a given Twitter bio contains four steps:
\begin{enumerate}
    \item {\it Split} the bio into separate phrases using a manually constructed list of explicit delimiters - As with \citet{yoder2020phans}, we iteratively construct a set of delimiters that we identified via manual inspection. The rest of the algorithm then proceeded on each phrase that was split out
    \item {\it Clean} each phrase using a set of manually defined substitutions, removals, and replacements - Again via manual inspection, we constructed a set of words, phrases, and \hl{symbols} that constituted noise that could be removed without changing the meaning of a phrase. This include punctuation, stopwords like ``to'' and ``from'', and the phrases ``i love'' and ``i am a''. 
    \item {\it Filter} out phrases that are unlikely to be personal identifiers - We found that these first two steps generated three classes of phrases that needed to be filtered out. First, because we used the period (.) as a delimiter, the algorithm extracted out several one-letter phrases. We also filter out two-letter phrases, which mostly consisted of difficult-to-interpret abbreviations.  Finally, as described below, we found empirically that phrases which were longer than 4 whitespace-delimited tokens were frequently not personal identifiers.
    \item {\it Return} the filtered, cleaned set of phrases extracted from the bio
\end{enumerate}

This algorithm implicitly provides answers to two questions. First, {\it what ``counts'' as a personal identifier?} The algorithm in general assumes that \emph{any} delimited phrase in a bio is a (potential) personal identifier, except for one and two-token expressions, higher-order ngrams, and the delimiters themselves. Further, the algorithm also only assumes phrases are separate if they use one of our delimiters, meaning that many descriptions end up consisting of one, long phrase. 

Second, the algorithm answers the question, {\it when is someone expressing a single identity versus multiple identities?} Questions about what constitutes a single identity have been addressed in a variety of ways. For example, Affect Control Theorists use the notion of modifier terms to reflect their belief that, for example, a ``doctor'' and a ``mean doctor'' represented references to the same identity, but with different affective meanings. Scholars of intersectionality \cite{crenshaw_mapping_1991} emphasize the problems caused by thinking of the compound phrase ``black woman'' as being the combination of two separate identities, black and woman.  Our algorithm relies on the user to express how they see their identities as being separated. So, for example, a profile stating ``I am a black woman'' would return a single personal identifier (``black woman''), whereas a profile stating ``I am black, and a woman'' would contain two personal identifiers (``black'' and ``woman'').  In our discussion section, we consider potential issues with this approach.

We also acknowledge one question the algorithm does \emph{not} address: {\bf when should we consider two individuals to have the same identity?} For example, does a ``father of 3'' have the same identity as a ``father of two''? Again, we leave further consideration of this point to our discussion.

Overall, then, our method makes little attempt to differentiate phrases that are likely personal identifiers from those that are not. We would, therefore, be surprised if all extracted phrases were personal identifiers, and as such we propose below a number of ways to validate the method's precision - that is, the percentage of extracted phrases that are actually personal identifiers.

\subsection{Methods Used for Validation}

\begin{table}[]
    \centering
    \begin{tabular}{|l|p{5.5cm}| p{2cm} | p{4.8cm} |} \hline
    & {\bf Evaluation} & {\bf Dataset(s)} & {\bf Purpose} \\ \hline
    1 & {\bf Manual annotation for different properties of personal identifiers} &  Panel 1 & Inter-rater Reliability, Evaluate Algorithm Precision \\ \hline
    2 &{\bf Comparison to demographics and status markers on Twitter}   & Panel 1 & Convergent Validity, Discriminant Validity \\ \hline
    3 & {\bf Clustering of personal identifiers} & Panel 1 & Convergent Validity, Utility \\ \hline
    4 &{\bf Manual annotation of probabilistic sample} &  Panel 1, Panel 2, Random &  Generalizability, Inter-rater Reliability,  Evaluate Algorithm Precision \\ \hline
    5 & {\bf Comparison to existing lists of identities} & Panel 1, Panel 2, Random & Generalizability, Convergent Validity, Discriminant Validity, Utility  \\ \hline
    \end{tabular}
    \caption{A summary of the five different analyses we perform to assess the generalizability, validity, and/or utility of our measure, along with the datasets considered.}
    \label{tab:eval}
\end{table}

Table~\ref{tab:eval} outlines the five different approaches we used to evaluate the validity, utility, and generalizability of our methodology. The first three evaluations are conducted only on Panel 1, the dataset on which our initial rounds of qualitative analysis were carried out. As with other work combining qualitative and qualitative analysis \cite{nelson_computational_2020,eads2020separating}, we expected that the proposed method wold work best on the data upon which our qualitative work was carried out. Nonetheless, it is instructive to assess the generalizability of our proposed approach to provide evidence of its utility, or lack thereof, beyond our data of interest. To this end, the final analyses are conducted on all three datasets of user bios described above. Below, we provide more detail on each of these \hl{analyses}.

\subsubsection{Manual annotation for different properties of personal identifiers}

Our first round of annotation focused on evaluating the odds that an extracted phrase was a personal identifier when controlling for 1) the number of whitespace-delimited tokens in the phrase and 2) the number of unique bios the phrase appeared in.  We also used it to evaluate the inter-rater reliability of the concept of personal identifiers.

To construct the annotation task, we first ran our algorithm on the Panel 1 dataset. We then split extracted phrases depending on whether they were \emph{unigrams} (1 token), \emph{bigrams} (2 tokens), \emph{trigrams} (3 tokens), or higher-order ngrams (4+ tokens). We also split phrases on whether they appeared in only 1 unique bio, only 2, 3-5, 5-10, 10-25, 25-100, or 100+. For each combination of token and unique bio count, we then sampled 30 identifiers, for a total of 840 identifiers to be annotated. All identifiers were sampled with equal probability.

Annotators were the three paper authors. Sampled phrases were given randomly to two authors each, and a third was used to resolve disagreements. Annotators were asked, for each phrase, ``Is this a personal identifier?'' Each phrase was originally labeled with one of four labels: ``Yes,'' ``No,'' ``Yes, but potentially more than 1 identifier is presented,'' and ``Unclear''. The Unclear label was used when annotators could not come to a definitive yes-or-no conclusion, and was used for all non-English terms, because we did not want to rely on translation software. The ``Yes, but...'' option was used sparingly, and was combined with the ``Yes'' option. Results are therefore reported with only three categories - Yes, No, and Unclear. The full annotation instruction document we constructed is provided in the Supplementary Material attached to the article.

To evaluate inter-rater reliability, we used Krippendorf's alpha \cite{krippendorffReliabilityContentAnalysis2004}. We provide reliability statistics both with and without phrases that annotators felt were sufficiently unclear that they could not be annotated accurately as Yes or No.  Data was coded twice to evaluate how discussion would increase inter-rater reliability.

\subsubsection{Comparison to Demographics and Status Markers on Twitter}

We use demographic information associated with voter-linked Twitter accounts in the Panel 1 dataset, as well as markers of status on Twitter, to evaluate the \emph{convergent validity} and \emph{discriminant validity} of our method.  Convergent validity encapsulates ``the extent to which [a] measure matches existing measures that it should match'' \cite[][pg.216]{quinn2010analyze}. In our case, we expect that identifiers which either explicitly (e.g. ``woman'') or denotatively (e.g. ``wife'')  signal demographics are correctly associated with that demographic.  Given that social identity is a narrower set of categories than demographics, we should also expect that it is able to capture narrower social groups that demographics are not intended to capture. While this overlaps with convergent validity, it is also a form of \emph{discriminant validity}, defined as ``the extent to which the measure departs from existing measures where it should depart'' \cite[][pg.216]{quinn2010analyze}.

With respect to demographics, we consider three variables that are categorical in our data--sex, political affiliation, and race, and two continuous variables, percentage of the user's county in a rural area, and age.  With respect to status variables on Twitter, we use the categorical variable that shows whether or not a user was verified, and a common continuous measure of status, the friend/follower ratio \cite{sylwester2015twitter,wang2010detecting}. This measure is computed as $\log((\# Friends + 1)/(\#Followers + 1))$.

Our approach to assessing convergent and discriminant validity given these variables has two steps. First, we compute a measure of the relationship between each personal identifier and each variable.  We then look at the top 10 personal identifiers most heavily aligned with different ends of the spectrum defined by each variable (e.g. for age, young versus old).  We look only at these more extreme relationships because we expect that most identifiers are not cleanly aligned with any given \hl{dimension of social meaning} \cite{joseph_when_2020}. \hl{Additionally, b}ecause these two forms of validity require a degree of subjectivity, we therefore believe they were best assessed by looking only for whether or not more extreme observations aligned with our expectations.

Where the demographic \hl{(or status variable}) is continuous, we simply calculate the mean value for each personal identifier for all users who express that identifier in their bio. We then extract the 10 highest and lowest mean values. For categorical variables, we compute the log-odds of an identifier being aligned with users with a particular categorical value. More specifically, we compute two log-odds scores. The first, used only for visualization purposes, is a \emph{raw} log-odds score, which has an intuitive meaning representing a multiplier on the number of times an identifier is used by one demographic category more so than another. Assuming a categorical variable that can only take on two values, A and B, this log-odds score is computed as $\log((A+1)/(B+1))$, where a one is added to avoid infinite or undefined values. The second score is a \emph{normalized} log-odds score, which is used to rank-order identifiers  in a way that accounts for how frequently a phrase occurs overall. Because significant additional notation is needed to describe the quantity, we refer the reader to Equation 22 in \cite{monroeFightinWordsLexical2008} for full details. Intuitively, however, the method simply gives slightly more extreme values to identifiers where more evidence exists for the same log-odds score (e.g. it would give a more extreme value when $A=100$ and $B=10$ than when $A=10$ and $B=1$, even though these have the same log-odds score, because there is more evidence for one over the other).  Note that because race is a multi-category construct, we follow prior work and compare how individuals assigned non-White race ethnicity values distinguish themselves from the culturally defined ``default identity'' of whiteness \cite{joseph_when_2020}.

To ensure privacy, all identifiers shown are associated with a minimum of ten unique bios.

\subsubsection{Clustering of Personal Identifiers}

\hl{In addition to studying the connection between personal identifiers and demographics/status indicators, we also study connections between the personal identifiers themselves. Specifically, we identify and analyze} clusters of personal identifiers that commonly co-occur within the same bios. To the extent these clusters align with traditional notions of how social identities are organized, these results can be used to show convergent validity. To the extent th\hl{ese clusters deviate from expectations, suggesting} future directions for the study of identity presentation online, we also use the clustering analysis to \hl{show} that our method has \emph{utility}. Utility is tightly intertwined with validity, and certain forms of validity, e.g. \emph{hypothesis validity}, are aimed specifically at utility \cite{jacobs2021measurement}. Here, we prefer the more general term utility because it best captures our intent.

Because we aim to use clustering to validate our method, and not as a novel methodology in and of itself, we try to adhere as closely as possible to a similar methodology used on similar data in prior work. Specifically, we use the methodology established in recent work from \citet{joseph2020says}. In their work, they use \emph{Spectral Co-Clustering}, a methodology originally developed to cluster documents together with words frequently used within them \cite{dhillonCoclusteringDocumentsWords2001}, as a way to study how hashtags cluster based on their co-usage by Twitter users.  They show that the method, which can be applied in the same situations as topic modeling, has several useful properties, \hl{including that it requires} only a single \hl{tuneable} parameter, which is the number of possibly clusters ($k$).

More formally, spectral co-clustering is a technique used to cluster a bipartite graph. A bipartite graph is, in turn, a matrix where the rows and columns have different semantics, and cells represent relationships between them. In \citet{joseph2020says}, the rows of their matrix are hashtags, the columns are users, and cells of the matrix denote how frequently a given user tweeted a given hashtag. The only difference in our case is that the rows are personal identifiers, not hashtags. Spectral co-clustering performs clustering by optimizing a relaxation of the min-cut problem, labeling each row and column with a cluster label such that the value of edges which span row/column pairs not in the same cluster are minimized, and edges within clusters are maximized. As with \citet{joseph2020says}, we leave a full mathematical derivation of the algorithm to the original work \cite{dhillonCoclusteringDocumentsWords2001}. Beyond changing our focus from hashtags to personal identifiers, we otherwise mimic the setup used by \citet{joseph2020says}, in that we use the same algorithm with the same fixed number of clusters, $k=100$.

The output of Spectral Co-clustering is a group of personal identifiers, and the users that tend to \hl{express} them in their bios. Because we opt to list all identifiers used in the analysis in the paper, we limit ourselves to the 2756 phrases extracted by our algorithm that appeared in more than 100 bios. Additionally, because users with only zero or one personal \hl{identifiers} do not provide information for clustering, we also limit our analysis to the 337,139 users who had more than one of these 2,756 identifiers in their profile. Finally,  we focus only on the clustering of personal identifiers, and not their overlap with specific users.

\subsubsection{Manual annotation of Probabilistic Sample}

Our second round of annotation focused on generalizability, as well as validity with respect to the overall proportion of extracted phrases that were personal identifiers. With respect to generalizability, we considered all three datasets of user bios. We sampled 200 identifiers from each dataset probabilistically, based on the number of bios in which an identifier occurred. So, for example, an identifier occurring in 100 bios was 100 times more likely to be sampled in this round than an identifier appearing in only one bio. The evaluation then proceeded exactly the same as in our first manual annotation described above, except that we did not perform a re-coding of the data after discussing disagreements.

\subsubsection{Comparison to Existing Lists of Identities}

Our final evaluation compares the phrases extracted by our method to six different existing lists of identities and/or other terms that have been previously used to study social identity.  Doing so provides a form of convergent validity (our method should capture at least some of these identities) and discriminant validity (our method should, according to our claims, also capture other ways in which bios present identifying information beyond identities). As a way of showing the utility of extracting out personal identifiers from Twitter bios, we also use survey data from prior work to explore differences in which \hl{identities} are and are not expressed in bios.

The first three lists we use are Identities (N=931), Behaviors (N=814), and Modifier terms (i.e. adjectives) (N=660) from Affect Control Theorists (ACT) \cite{Smith-LovinInterpretingRespondingEvents2015}. All phrases in these lists are tied to survey data that characterize the term along three dimensions of affective meaning - Evaluation (how good or bad the identity, behavior, or modifier is perceived to be), Power (how powerful or powerless it is), and Activity (how active or passive it is). These meanings are determined via a widely validated survey instrument \cite{heise_expressive_2007}.  

The last three lists are sets of identities used to study stereotyping in recent literature in natural language processing. \citet{bolukbasi_man_2016} use a set of identities (N=320), mostly occupations, to study gender bias in word embeddings. These identities are linked to perceived Gender using survey methods like those in ACT \cite{Smith-LovinInterpretingRespondingEvents2015}. \citet{agarwal_word_2019} use a set of identities (N=373), including national identities and occupations, to study how word embeddings capture stereotypes of personality traits. These identities are provided with survey-based perceptions of their position along the Big 5 personality traits. Finally, we use a list of 310 identities from a recent CSCW paper studying expressions of identity  on Twitter \citet{joseph_girls_2017}. These identities are not attached to any meaning measurements.

All lists of identities, and their associated survey data, are included in the Supplemental Material provided along with this article.

\section{Inter-rater Reliability}\label{sec:reliability}

On our first round of annotations (Evaluation 1 in Table~\ref{tab:eval}), Krippendorf's alpha was .57, and .69 when excluding phrases one or more annotators marked as unclear. After discussion and re-annotation of the same terms, these values rose to .71 and .81, respectively, reflecting moderate to substantial agreement on the concept of a personal identifier amongst annotators. We observed similar, moderate levels of inter-rater reliability in our second round of \hl{annotation} (Evaluation 4 in Table~\ref{tab:eval}). Overall, Krippendorf's alpha was .65 for all datasets, and .72 when excluding phrases marked as unclear by at least one annotator. The overall number was .63, .66, and .66, and was .72, .74, and .69 when excluding the Unclear responses for Panel 1, Panel 2, and the Random dataset, respectively. 

While guidelines on necessary levels of reliability are notoriously subjective \cite{mcdonald2019reliability}, these reliability levels are similar to those for other developing constructs in the CSCW literature, with the closest we found being hate speech \cite{elsherief2018hate} (0.622). These levels of reliability therefore suggest that while challenges remain in solidifying the concept, personal identifiers are a reasonably salient and meaningful construct that can be used to address downstream research questions.  Output from all three formal rounds of qualitative coding are provided in the Supplementary Materials submitted with this article.

\section{Validation Results}
\subsection{Overview}

In the following sections, we provide results for the validation analyses described above. In order to provide context for this, however, we give here some initial descriptive statistics. \hl{These statistics} also provide a degree of initial face validity for the method for the Panel data, and indicate caveats to generalizing to a random sample:
\begin{itemize}
    \item  We extracted 57,427 unique unigrams, bigrams, and trigrams from Panel 1, 57,292 from Panel 2, and 38,004 from Random. Of these, 25,496, 24,981 and 5,715 were observed in more than ten unique bios in Panel 1, Panel 2, and Random, respectively.  There was a correlation of .83 for the number of unique bios \hl{that} phrases occurred in between the Panel 1 and Panel 2 datasets. This correlation was .34 between Panel 1 and Random.
    \item We extracted at least one \hl{phrase that our algorithm defined as a} personal identifier from 71.0\% of the bios in Panel 1, 70.8\% of bios in Panel 2, and 53.2\% of bios in the Random sample.
    \item The top 5 extracted phrases in both Panel 1 and Panel 2 were wife, husband, writer, mom, and father; at least one of these terms appeared in 6.4\% and 6.3\% of all bios in these two datasets. In contrast, the top five extracted phrases in the Random dataset were instagram, music, writer, wife, and love. Notably, only three of these five are personal identifiers.
\end{itemize}

\subsection{Manual annotation across token and bio count}

\begin{figure}[t]
	\centering
	\includegraphics[width=\textwidth]{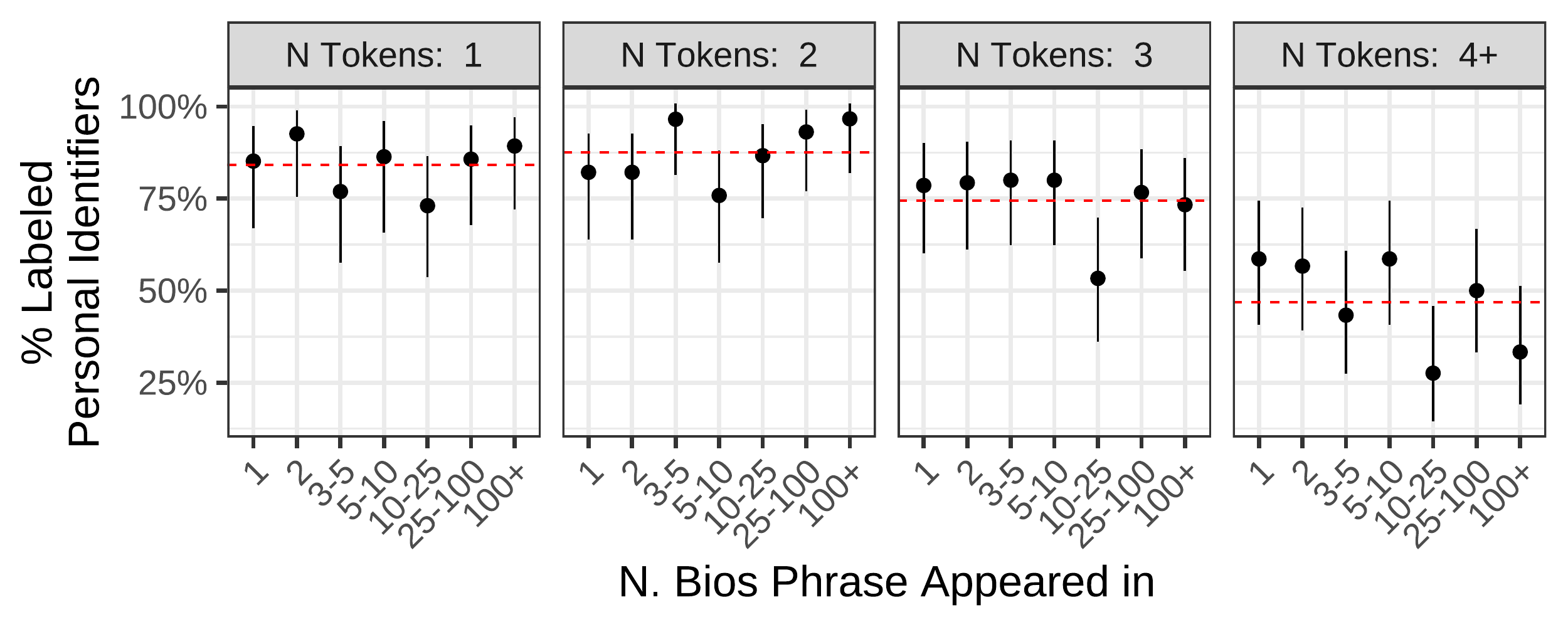} 
	\caption{For a given set of phrases with a specific number of tokens (separate sub-plots) that appeared in a specific number of user bios (x-axis), the estimated percentage of \hl{those} phrases that are personal identifiers. \hl{Results are} based on our manual annotation. Each error bar shows a mean estimate and confidence interval (estimated using the Agresti-Coulli method \cite{agrestiApproximateBetterExact1998} for computing binomial confidence intervals). Red bars indicate the average across all phrases annotated with a given number of tokens, irrespective of the number of bios appeared in. 
}
	\label{fig:ann_1}
\end{figure}

We find that unigrams, bigrams, and trigrams extracted from bios in the Panel 1 dataset were likely to be personal identifiers, but higher-order ngrams were not. We also find that this was true irrespective of the unique number of bios the phrase appeared in.
Results are presented graphically in Figure~\ref{fig:ann_1}. The figure shows that there was no consistent relationship between the number of unique bios a phrase appeared in and the odds of it being a personal identifier. The one potential exception to this were trigrams that appeared in 10-25 different bios. These were significantly less likely to be personal identifiers than in any other subset of trigrams. However, a re-review of all trigrams did not provide any clear reason why \hl{phrases} appearing in this range of bios were distinct from other trigrams, and the pattern is not part of any trend in the graph. We therefore do not believe there to be any clear, systemic underlying cause for this result.

Averaging over the number of bios a phrase appeared in, we find that 84\%, 88\%, 74\%, and 47\% of extracted unigrams, bigrams, trigrams, and higher-order ngrams were personal identifiers, respectively.  Results therefore suggest a need to focus only on extracted phrases containing a small number of whitespace delimited tokens.  As noted above, we therefore opted in our final algorithm, used for all other analyses presented below, to filter out phrases with more than 3 whitespace delimited tokens.

\subsection{Comparison with Demographics and Twitter Status Measures}

\begin{figure}
	\centering
	\includegraphics[width=\textwidth]{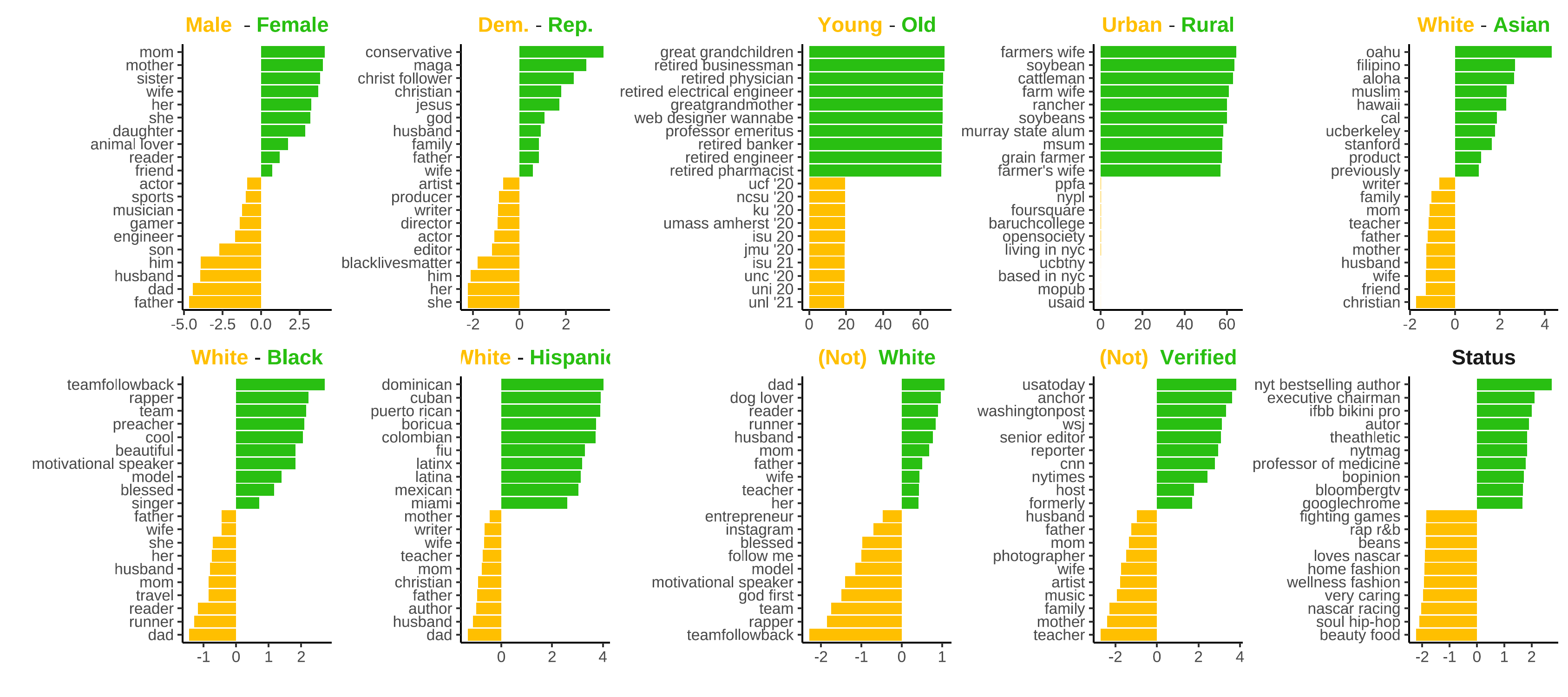}
	\caption{Each subplot presents the top ten identities in Panel 1 associated with different ends of various demographic categories and/or Twitter status variables. Where the demographic is categorical, identities shown are those that have the highest normalized log-odds. We present the raw log-odds value \hl{in the figure}, however, because it has an intuitive meaning - for example, the top, leftmost subplot shows that the identifier ``father'' is almost four times as likely to be seen in a user's bio if the user stated they were a male on their voter record, and the identity mother is almost three times more likely to be seen in a user's bio if they stated they were female in their voter record. This is displayed by linking the title of the plot to the color and direction of the effect. For example, in the top left, Male is colored orange, and to the left of Female. Thus, bars that are orange and stretching to the left reflect more male-oriented personal identifiers (in our data).  Where the variable is continuous (Status, Urban/Rural, Young/Old), we take the identities with the highest (\hl{or} lowest) mean values across all users expressing that identity. The orange bars in this case are simply much lower on the relevant scale than the green bars. We show only identifiers appearing in more than ten bios.
}
	\label{fig:identity_demog}
\end{figure}

Our analysis of the personal identifiers that are most highly correlated with demographics and Twitter status variables provide evidence of convergent and discriminant validity.  Results of this analysis are displayed in Figure~\ref{fig:identity_demog}, which shows the top ten identities most heavily aligned with each end of the demographic categories we constructed.\footnote{In the Supplementary Material, we provide results for the top 50 personal identifiers most correlated with each dimension. General claims made about validity based on Figure~\ref{fig:identity_demog} hold for those presented identities as well.}

With respect to convergent validity, the most salient examples are along dimensions of gender, age, urban/rural, Hispanic ethnicity, and verified status. With respect to gender, denotatively gendered terms represent the top five most male and most female words. This is also true of age, where all but one of the oldest identifiers refer to being a great grandparent or being retired, and all ten of the youngest identifiers are college graduation dates that would imply the individual is approximately 21 years old. With respect to the urban/rural divide, all but one of the personal identifiers associated with users in urban areas directly reference living in New York City, or an organization or college based in the city. The ten identifiers most strongly correlated with living in a rural area are all related to farming or to Murray State, a university in rural Kentucky. With respect to Hispanic ethnicity, all but one of the top-ranking terms refer to locations within or outside of the United States with large Hispanic communities, or to alternative labels for individuals presenting Hispanic identity (e.g. Latinx). Finally, identifiers in verified profiles indicate membership in news media, one of the most dominant groups of verified users \cite{paul2019elites}.

With respect to discriminant validity, we again highlight dimensions of gender and Hispanic ethnicity, as well as partisanship and the Asian/White dimension, as cases where our method diverges in valid ways from the aligned demographic variable. Along gender, we see that women are 1.66 times less likely to identify as engineers, 1.37 times less likely to \hl{claim the identity} gamer, and 1.78 times as likely to present the identifier animal lover. These choices on how to present the gendered self represent established structural constraints in tech \cite{shih2006circumventing} and normative constraints of woman as empathizers and men as systematizers \cite{bianGenderStereotypesIntellectual2017}. With respect to partisanship, we see an expected (convergent) relationship between the terms conservative and being a Republican. However, we also see links to racial justice and the expression of pronouns for personal identifiers used most by Democrats, and to family and religion for Republican personal identifiers. Finally, we see that personal identifiers most often used by Asian and Hispanic people, relative to White people, provide a more detailed view on the more fine-grained racial categories that are summed into the larger demographic categories.  These observations all align with the expected ways in which personal identifiers, as more narrow categories than demographics, can capture more fine-grained and \emph{relevant} information about individuals, e.g. more detailed racial categories, and more nuanced representations of gender \cite{ridgeway1999gender} and partisanship \cite{dellapostaWhyLiberalsDrink2015} as systems of cultural meanings. This provides evidence of discriminant validity. 

Finally, Figure~\ref{fig:identity_demog} shows that it is important to differentiate personal identifiers, which are what individuals choose to express, from demographics, which are assigned (albeit often problematically \cite{hamilton_genealogy_2020,haimson_constructing_2016,pennerEngenderingRacialPerceptions2013}) to individuals. This is clear in that Figure~\ref{fig:identity_demog} shows that there were no identifiers that denotatively provided markers of Whiteness along any of the racial dimensions we studied.  This is likely due to the well-established fact that White people consider their race to be the ``default'' in America, and thus rarely mark themselves with it \cite{mcintosh1988white}. 

To the extent that the identifiers are denotatively aligned with certain demographics, Figure~\ref{fig:identity_demog} therefore provides evidence of convergent validity. To the extent that personal identifiers can provide more fine-grained information than demographics can, but in ways that are still aligned as expected with certain demographics, this provides evidence of discriminant validity. Finally, Figure~\ref{fig:identity_demog}  emphasizes that  personal identifiers are not a substitute for demographics, but are instead a complementary form of analysis. Specifically, studying personal identifiers in social media bios in concert with demographic analyses can help us to better understand how individuals with different demographics may be constrained in the identities that they are able to access or choose to present.



\subsection{Cluster Analysis}

\begin{table}[]
    \centering
    \begin{tabular}{p{.1\textwidth}|p{.8\textwidth}}
{\bf Clust Num.} & {\bf Associated Personal Identifiers} \\ \hline
  3 & assistant professor, associate professor, chair, faculty, highered, opinions my own, phd candidate, tweets my own, views mine, yale \\ \hline
79 & believer, brother, chaplain, christ follower,  church planter, disciple, disciple of christ, encourager, follower, follower of christ, follower of jesus \\ \hline
35 & ally, biden2020, blm, bluewave, democrat, disabled, fbr, lgbtq, liberal, metoo, nevertrump, resist \\ \hline
73 & backtheblue, conservative, constitution, god bless america, kag, maga, my country, nra, patriot, prolife, qanon, trump supporter, wwg1wga \\ \hline
22 & adrenaline junkie, amateur photographer, animal advocate, avid traveler, chicago native, chocolate lover, coffee snob, dog enthusiast, dream chaser, fashionista, fitness fanatic, fitness junkie, food blogger, marketing guru \\ \hline
63 &  auntie, aunt, child of god, cousin, daughter, friend, girlfriend, grandmother, homeschooler, jesus lover, mama, mimi, mom, momma, mommy, mother, nana, niece, nurse, pastor's wife, sister, stepmom, wife, wifey \\ \hline
87 & cat lady, cat mom, cat person, daydreamer, empath, enfj, ginger, grad student, gryffindor, infj, infp, intj, introvert, mama bear, mental health advocate, old soul,pastry chef, social worker\\ \hline
38 & always, also, all, ask me, bad, best, but, cheers, don't worry, dude, end, ever, every day, everything \\ \hline
    \end{tabular}
    \caption{A sample of personal identifiers from eight of the clusters identified by the spectral co-clustering we ran; information on all identities for all clusters is presented in the Appendix in Table~\ref{tab:app_clust}. We refer to the (arbitrarily assigned) cluster number in these tables in order to organize our discussion.}
    \label{tab:main_clust}
\end{table}

We find that personal identifiers cluster into collections of identities that 1) we would expect to see together in Twitter bios (convergent validity), 2) suggest new insights for future analysis (utility), and 3) belie noise in our extraction process.  We use the sample of clusters in Table~\ref{tab:main_clust} to discuss examples of each of these types of clusters. As in \citet{joseph2020says}, we do not claim that this list is exhaustive of all patterns in our clustering, but rather use them to explore qualitatively the patterns brought to light.

The first type of cluster, which show convergent validity, are those that align with social institutions - collections of identities that can be organized with respect to a particular institutional setting \citet{heise_self_2010-1}.   Clusters 3 and 79 provide examples of identifiers organized around the institutions of Higher Education and the Church, respectively. Clusters 35 and 73 are representative of the institution of Politics, although as one would expect, they are separated into left- and right-leaning identifiers.  Institutionally-aligned clusters of identities have emerged in prior studies that look at how people talk \emph{about} social identities \cite{heise_self_2010-1,joseph_exploring_2016}, we are the first we are aware of to observe this in how people talk about themselves.

The second type of cluster, which suggest interesting avenues for more detailed study, are represented by Clusters 22, 63 and 87. These clusters each presented collections of identifiers that displayed interesting ways in which users navigated the context collapse. Rather than present an identity aligning with a single institution (e.g. politics), users in Cluster 22 used identifiers across institutional boundaries that seemed to present a more active, exciting, cultured, and talented self. Given the paucity of data on the ways in which individuals seek various identities across social institutions, Cluster 22 shows that social media bios can be a rich new source of analysis for questions in social psychology that target why and how individuals select out specific combinations of identities when composing a self image \cite{ramarajanPresentFutureResearch2014}.  Clusters 63 and 87 also are relevant to this question, but are specifically interesting in that they show how users present gender differently. Both clusters contain terms denotatively aligned with being a woman, but are otherwise dissimilar. In Cluster 63, gender is presented in ways tightly aligned to religion and family; in 87, as a component of a more specific identity (cat lady) and as a component of other definitions of the self, especially personality types (infj, infp, etc.). These clusters \hl{again} point to the potential use of identifiers in bios to continue to shape our understanding of how distinct perceptions of gender can arise in distinct social contexts \cite{ridgeway_framed_2011}.

The final type of cluster, represented by Cluster 38, represents collections of phrases that were almost exclusively not personal identifiers. As with \citet{joseph2020says}, these uninteresting clusters are useful in that they provide \hl{a mechanism} to rapidly filter large datasets down to subsets that are of interest for further analysis. 

\subsection{Manual Annotation of Probabilistic Samples}

\begin{figure}[t]
	\centering
	\includegraphics[width=.8\textwidth]{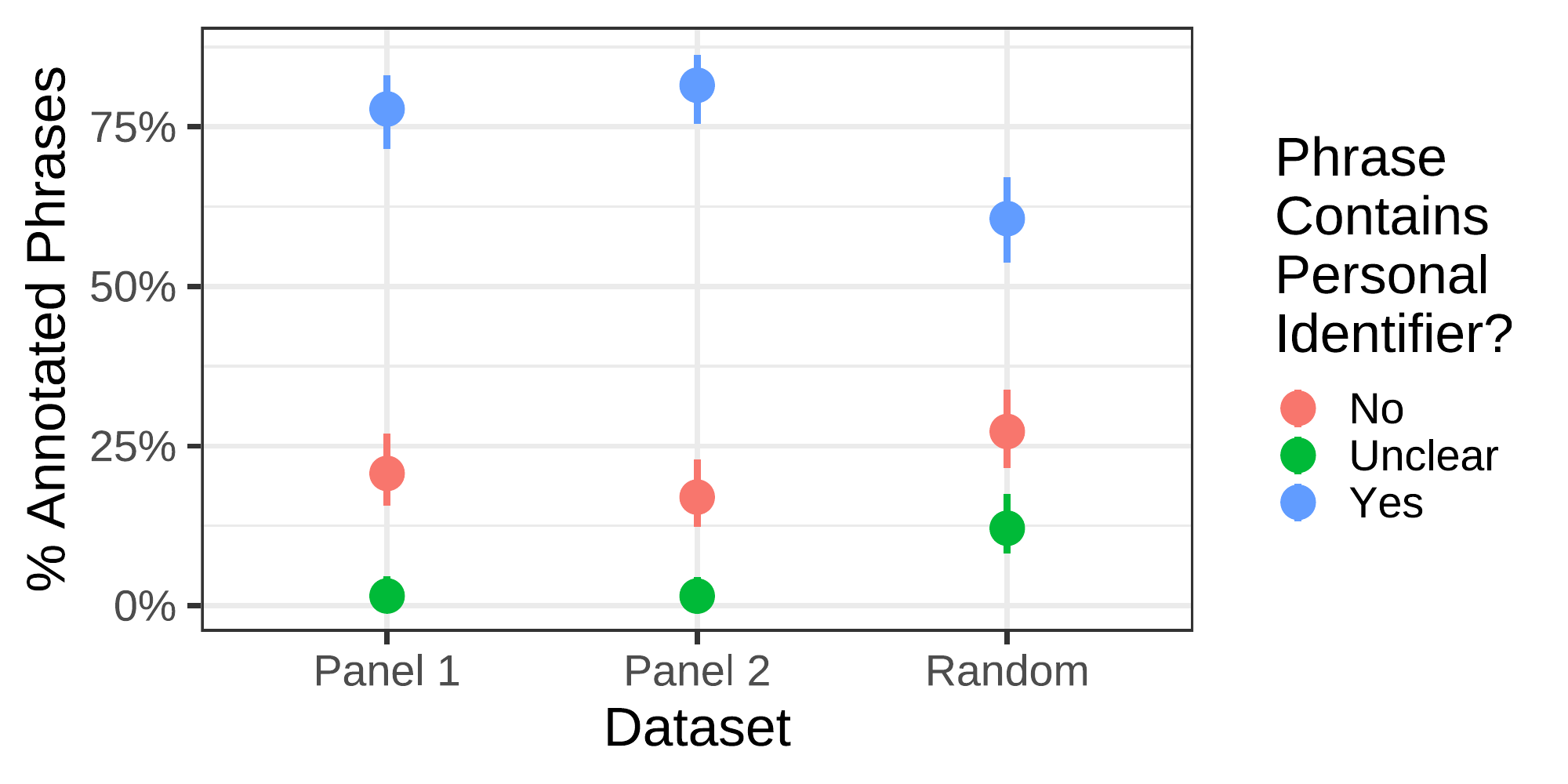} 
	\caption{For a given dataset (x-axis), the estimated percentage of phrases that are personal identifiers (blue), are not personal identifiers (red), or that were annotated as unclear (green) based on our manual annotation. Each error bar shows a mean estimate and confidence interval (estimated using the Agresti-Coulli method \cite{agrestiApproximateBetterExact1998} for computing binomial confidence intervals).
}
	\label{fig:ann_2}
\end{figure}

Figure~\ref{fig:ann_2} shows that extraction of personal identifiers from bios is reliable for accounts linked to voter registration records, but less so for extraction from randomly sampled accounts. Of the 200 phrases sampled for the Panel 1 and Panel 2 datasets,  77.8\% and 81.5\% were labeled as personal identifiers, 20.7\% and 17.\% were labeled as not personal identifiers, and only 1.5\% in both cases were unclear. In contrast, only 60.6\% of phrases extracted from the Random dataset were personal identifiers, compared to 27.3\% not personal identifiers, and 12.1\% labeled as unclear. 

Extracted phrases that annotators agreed were not personal identifiers in Panel 1 and Panel 2 fell largely into one of two categories - phrases that could not be interpreted without additional context (e.g., ``formerly'' and \hl{``contributor to''}) or that were overly generic (e.g. ``all comments,'' ``is my life,'' or ``laughing''). Phrases marked as unclear were either ambiguous acronyms/hashtags (\#plf), or due to polysemy (it was unclear, out of context, if ``trusts'' referred to a trust or to the verb form of the word). In addition to examples of these \hl{issues}, a significant portion of the phrases marked Unclear in the Random dataset (18 of 24) were marked so because of the explicit coding decision to label non-English terms as Unclear. Use of translation software suggests that many of these were indeed identifiers (e.g. ``ingeniero de sistemas'' and ``17 anos''), but we opted to remain with our pre-determined coding rule and to therefore not re-annotate these after translation.
	
\subsection{Comparison with Existing Lists}
	
	\begin{figure}[t]
		\centering
	     \begin{tabular}{c c}
		      \includegraphics[width=.5\textwidth]{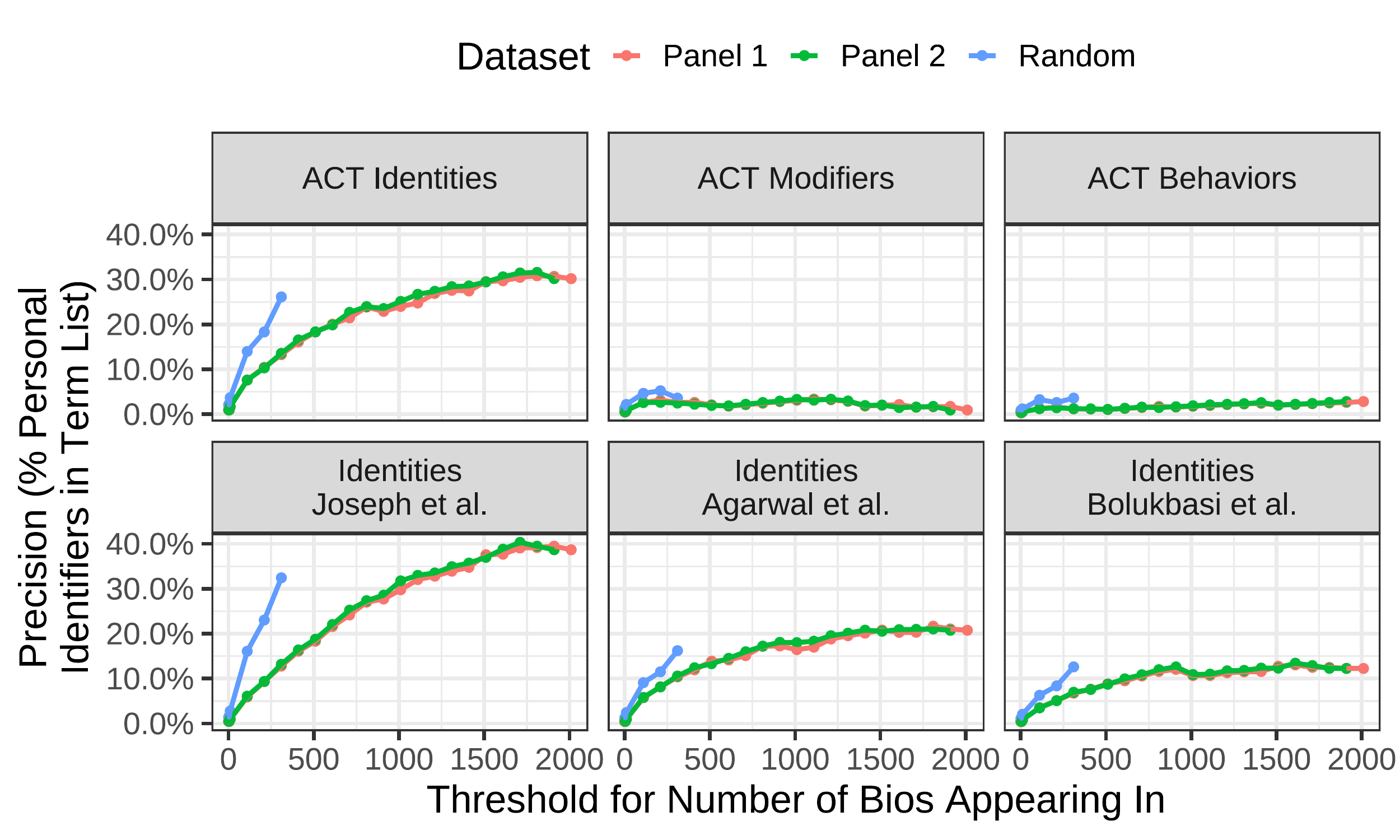}   & 
		      \includegraphics[width=.5\textwidth]{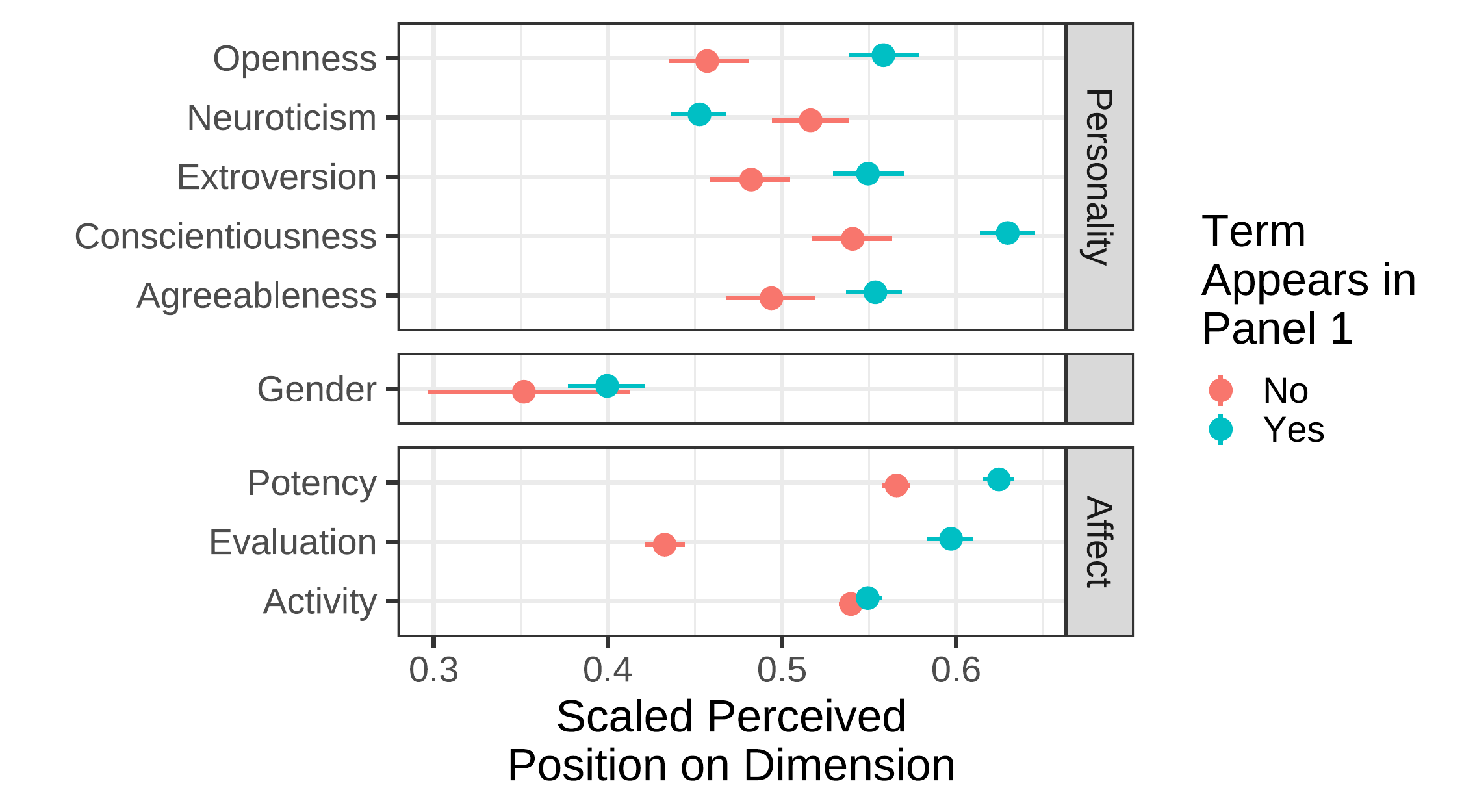} \\
		         (a) & (b)
		    \end{tabular}
		
		\caption{A) The x-axis presents a threshold, where we consider only identifiers appearing in greater than that number of unique bios. The y-axis shows, of the \hl{remaining} identifiers, what percentage were in a given pre-existing list (the lists are differentiated by the separate subplots). We show results for the different user bio datasets in different colors. We only show results for thresholds for which the set of identifiers left is greater than 100 to avoid instability.
		B) The y-axis presents different dimensions of meaning along which prior work provides \hl{measurements of} perceptions of identities in a pre-defined list. The y-axis presents the scaled and centered (to 0.5) mean value (with 95\% bootstrapped confidence intervals) of terms for which we have measurements on that meaning dimension. We differentiate this measure by whether or not the term did (cyan) or did not (red) ever appear as an extracted bio in the Panel 1 dataset.}
		\label{fig:lists}
	\end{figure}

We find that personal identifiers extracted from Twitter bios have significant overlaps with pre-existing lists of identities, but not behaviors or modifiers (convergent validity), and that they extend these lists in important ways (discriminant validity). Figure~\ref{fig:lists}a) shows that as we consider the set of personal identifiers in each dataset appearing more than N times (the x-axis), a larger percentage (y-axis) of those identifiers are in the various lists of identities considered. However, this relationship is \hl{non-existent} for modifiers and behaviors.  The figure shows results for each dataset until the number of remaining personal identifiers appearing in more than the given number of bios falls below 100. So, for example, only around 100 identifiers occurred in more than 300 bios in the Random dataset, and hence results do not extend \hl{beyond this number} along the x-axis.

Figure~\ref{fig:lists}a) shows, for example, that when considering only identifiers in more than 1000 bios in the Panel datasets, around 30\% of the\hl{m} can be found in the list of identities from the work by \citet{joseph_girls_2017} studying Twitter, and around 17\%, 25\%, and 10\% of them are in the lists used in \citet{agarwal_word_2019}, \citet{Smith-LovinInterpretingRespondingEvents2015}, and \citet{bolukbasi_man_2016}, respectively.  In contrast, only 2\% and 3\% of the extracted phrases are in the ACT list of behaviors and modifiers, respectively.  Even in the Random dataset, 25\% of extracted phrases appearing more than 200 bios are in the manually constructed list of identities used by \citet{joseph_girls_2017}. The fact that prevalent personal identifiers are in these existing lists of identities, but \emph{not} lists of general behaviors or modifiers, provides convergent validity. This is because it shows that our method extracts phrases that others have included in their manually constructed lists of identities.

There are, however, two remaining questions. First, what are the terms that are \emph{not} in any of these lists that \emph{are} personal identifiers? To address this, we looked at the top 25 phrases across all datasets that did not appear in any of the lists considered, as ranked by the number of bios the phrase appeared in. Notably included in this set were phrases that were not personal identifiers, like instagram and alum (the latter lacks context). However, also included were the phrases foodie, song writer, animal lover, photography, and politics. Each of these is a personal identifier, and provides useful information about the social identities \hl{that} a bio signals. We therefore find ample evidence of our method's utility relative to simply extracting phrases in fixed lists of identities. 

Second, what differentiates terms that \emph{are} in these pre-defined lists and that are \emph{not} in Twitter bios? Figure~\ref{fig:lists}b) addresses this question by plotting differences in meanings associated with identities that did (cyan error bars) or did not (red error bars) appear in at least one bio in the Panel 1 dataset. We find, using the measurements from \citet{agarwal_word_2019}, that identities Twitter users in Panel 1 chose to present at least once were perceived to be significantly more Open, Extroverted, Conscientious, and Agreeable, and less Neurotic. There was no significant difference in the presented identities in terms of perceived gender using the measurements from \citet{bolukbasi_man_2016}. Finally, presented identities, behaviors, and modifiers were found to be more powerful, and more good, but no more active or passive, using the data from \citet{Smith-LovinInterpretingRespondingEvents2015}.  In the Appendix, Figure~\ref{fig:lists_appendix}, we show results for the Panel 2 and Random datasets, which are consistent with these observations except along gender, where we observed that identifiers were significantly more likely to be identifiers perceived as being associated with men. Taken together, these results are most important in that they provide an example of the utility of linking social psychological measures of identity to self presentations on Twitter. In doing so, we gain a new quantitative method to study how people decide what aspects of the self to present in Twitter bios.

A final note pertains to an apparent discrepancy between Figure~\ref{fig:lists}a), which shows that phrases are more likely to be in lists of identities if they occur more frequently, and Figure~\ref{fig:ann_1}, which suggests that the number of unique bios in which a phrase appears is not related to the odds that it is a personal identifier. This can be explained by the fact, evidenced by the most popular identifiers in the Panel datasets,  that identities are more common expressions in bios in comparison to other forms of personal identifiers (e.g. preferences). This explains why Figure~\ref{fig:ann_1}, which considers all forms of personal identifiers, contrasts in this way with Figure~\ref{fig:lists}a), where only identities are relevant.




\section{Discussion}

Overall, \hl{the analyses we present} provide significant evidence that personal identifiers, and our method for extracting them, are a useful, valid, reliable concept and tool for the study of social identity in Twitter bios. However, we here discuss a number of important caveats and considerations to be taken into account in the context of these results.

\subsection{Culture, Meaning, and Shared Identity}
The most salient issue we struggled to resolve over the course of our qualitative analysis was that annotators sometimes needed to rely on intricate personal knowledge to identify a personal identifier. One example of this form of disagreement was the phrase ``bleed maize and blue''. This phrase is a personal identifier, representing a common expression of fans and/or alumni of the University of Michigan. To individuals familiar with this university, the phrase was easy to annotate as a personal identifier (an affiliation); to those not, it was ``easily'' not a personal identifier (essentially, it was gibberish).  

More broadly, this speaks to the fact that there will ultimately be difficult-to-capture differences in when a given word or phrase in a bio will be seen by a given person as a marker of identity, and when it will not be. Similar concerns can be raised about the interpretation of the \emph{meaning} of that identity, even independent of whether or not it is seen as marking an identity. For example, the personal identifier ``Living in Portland'' will mean something different to individuals living in Maine as opposed to those living in Oregon, despite both likely recognizing it as a marker of identity. How to address these kinds of cultural differences in denoting identity markers and their meaning is an interesting and important question for future work.

As noted above, a similarly difficult question that the present work does not address is,  when should we consider two individuals to have the same identity?  Is a ``mom'' the same as a ``mother,'' for example, even though we would not expect an individual to use both of them in the same bio? This issue arose in our clustering analysis, for example, when we saw that ``father of two'' and ``father of 2'' ended up in different clusters. Addressing this is important for future analyses tying to, for example, the well-worn theory of homophily, which requires that we are able to determine whether or not two people share an identity \cite{mcpherson_birds_2001}. 

\subsection{Methodological Limitations}

The above points will likely require additional theoretical and empirical work alongside any methodological development. In contrast, there are two limitations of our existing algorithm for extracting personal identifiers that may perhaps be addressed by methodological advances alone. Specifically, the method we develop 1) performs limited filtering and 2) does not extract sub-clauses of phrases. We opted for these more straightforward options here because we believed the analysis was complex enough without the need for additional methodological considerations, e.g. the use of machine learning. 

We expect that future work which filters or subsets more intelligently could improve on extraction. To this end, our work provides annotated data to evaluate the precision of such algorithms. However, it is notable that we do not provide data to evaluate the recall of our method. That is, we do not provide any analysis of the number of personal identifiers our approach does \emph{not} capture. This is because of the unresolved issues noted above about the similarity of two identities. For example, is a model that extracts the phrase ``father'' from a bio with the statement ``father of two'' right, or wrong? Addressing recall therefore required answering a theoretical question that was beyond the scope of the present work.

\subsection{Generalizability}

There are three important considerations to be noted with respect to the generalizability of the present work.   First, throughout this paper, we have been careful to emphasize that we are studying \emph{self-presentation}, i.e. the process by which individuals opt to perform various identities within the specific context of a Twitter bio.  As discussed above, then, our results should be bracketed as an analysis only of the self as presented on Twitter, encompassing only individuals who self-select into Twitter and only how they choose to present themselves in their bios specifically.  Second, and related, it is not clear how well our methods or findings will generalize to social media bios on other platforms. \hl{On the one hand}, we believe there are a sufficiently large amount of reasons to be interested in self-presentation on Twitter specifically.  \hl{On the other,} there are long-established dangers of Twitter as the ``model organism'' \cite{tufekci_big_2014}, and we are therefore careful to speculate about cross-platform generalizability before we have conducted empirical work. To this end, we believe that the most related social media sites for this analyses appear to be tumblr, given the work of \citet{yoder2020phans}.

Finally, and most obviously, our validation analyses make clear that the method did not perform nearly as well on a random sample of Twitter users as it did for users linked to voter records.  We believe that there are at least two major causes for this. First, we found that there were far more non-English descriptions (even after filtering out based on last tweet) in the Random sample. Second, it generally appeared that the voter-linked users, who by definition of the dataset presented their real identity, seemed simply to use Twitter in different ways than the Random sample. Anecdotally, their use seemed to be more professional, which correlated with more active accounts, more complete bios, and wider usage of delimiters to separate multiple identities. To this end, we see the relative lack of generalizability of our method to the random sample also as a potential opportunity to better understand why certain individuals present themselves primarily with personal identifiers, while other seem not to. Given the correlation with more professional, more active accounts, this may lead to interesting new perspectives on who shapes discussion on Twitter, and in which corners of the platform this is true.

\subsection{Privacy and Misuse}

As we discuss above, the present work uses highly sensitive data on voter registration records that we are not legally allowed to share. However, it is also important to reflect on other questions about privacy and potential misuse of the proposed methods.

First, with respect to releasing data, we considered at length the relative risks to individuals in our datasets as compared to the merits for open science. We determined that the most effective balance was a two-pronged strategy. First, our data release contains all data points used to construct the plots presented here, as well as all annotated data from annotations for which reliability was calculated. Second, we will release, upon request to the last author, the Random dataset used here for replication purposes. We believe this approach mitigates potential issues with data privacy, while respecting the need for other scientists to replicate our work within the bounds of existing legal agreements.

Second, with respect to releasing our code, we believe that all code used here can be released responsibly to other academics and the broader public. We have two reasons for this decision. First, while we cannot necessarily control how others use our software, we have emphasized repeatedly here that using our tools to infer who someone ``is'' instead of the identities they are presenting will lead to invalid conclusions. As such, we hope that our work can serve to push  towards analyses where we label users as they state they would like to be labeled, rather than to try to infer some inherent truth about them. Second, and in a more sober sense, our method does not draw any new inference about individuals outside of what is presented in their profiles. As such, we do not see a clear way in which a malicious actor who already planned to use Twitter bios could do so in a way that our method would greatly extend.

Finally, however, we acknowledge that these decisions are driven by our own perspectives and life experiences. As such, requests for changes to these policies from individuals with other lived experiences will be seriously considered.

\section{Conclusion}

The present work proposes a new \hl{concept to characterize} how social identity is expressed in language, which we call the personal identifier. We provide a methodology to extract personal identifiers from Twitter bios, and rigorously evaluate the reliability, validity, generalizability, and utility of that method.  Overall, we believe we show convincingly that personal identifiers in Twitter bios can be used to study how users present themselves in new and useful ways.  As discussed above, the present work is not without limitation. However, we hope that our method, and its public availability, will encourage future mixed methods analyses that dig further into the structure of online identity as presented in bios and how it is constrained by various factors, both on and off platforms.  

\bibliographystyle{ACM-Reference-Format}
\bibliography{acmart}

\appendix

\section{All Clusters from Spectral Co-clustering}

\begin{longtable}{p{.1\textwidth}|p{.8\textwidth}}
{\bf Clust Num.} & {\bf Associated Personal Identifiers} \\ \hline
0 & 2 boys, 2 dogs, 2 cats, 2 kids, 3 kids, 4 kids, boy, college graduate, divorced, german, god is great, happily married, hubby, lds, living the dream, loves to travel, loving it, married, native american, retired, self employed \\ \hline
1 & 20 years old, 19 years old, 21 years old, ask, awesome, back, chill, employed, enough said, follow back, follow me, followback, followme, godfirst, i'll follow back, im me, imperfection is beauty, junior, just ask, keep calm, livin life, lol, music is life, single, taken, teamfollowback, teamiphone, teamleo, teamsingle, wild, xoxo, yolo \\ \hline
2 & 2011, 2012, 2013, 2014, 2015, 2016, 2017, 2018, 2019, 2020, addict, admin, aka, asst, asu, b a, b s, bilingual, bio, candidate, champion, class of 2017, cmu, coordinator, criminal justice, director of marketing, english, enthusiast, future, green, gvsu, j d, junkie, latina, m a, m s, msu, msw, nyu, ohio state university, phil, political, political science, present, pro, sociology, spanish, state, t1d, tamu, texas ex, txst, ucf, univ, usf \\ \hline
3 & 2021, assistant professor, assoc, associate professor, chair, dean, dir, equity, faculty, higher ed, highered, housing, inclusion, labor, mph, opinions my own, ph d, phd, phd candidate, phd student, prof, program director, public health, rts not endorsements, sapro, scicomm, tweets mine, tweets my own, views mine, views my own, yale \\ \hline
4 & 25 years old, 24 years old, 23 years old, 22 years old, baby, big, boys, cat, country girl, dig it, fam, first, foremost, irish, just a girl, just a guy, just living life, life is beautiful, live for today, live laugh love, marilyn monroe, nursing, oh yeah, old, period, proud of it, remember, sorry, that is all, that's about it, unt \\ \hline
5 & 4th grade teacher, 2nd grade teacher, 3rd grade teacher, 1st grade teacher, blessed wife, boy mom, boymom, dance teacher, dental hygienist, disney lover, elementary teacher, english teacher, fianc, first grade teacher, former teacher, girl mom, happy wife, hockey mom, kindergarten teacher, lover of christ, middle school teacher, mom of 2, mom of 3, mom of three, mom of twins, mom of two, mom to, mom to 3, mother of 2, mrs, preschool teacher, proud wife, school counselor, slp, special education teacher, travel agent, wife to \\ \hline
6 & abc, binders, bylines, cnn, columbiajourn, espn, forbes, huffpost, journalist, latimes, managing editor, medillschool, nbc, newhousesu, npr, nytimes, politico, reporter, reporter for, senior editor, time, usatoday, vice, washingtonpost, wsj \\ \hline
7 & accountant, business owner, cpa, dentist, entreprenuer, financial advisor, insurance agent, internet marketer, operator, property manager, real estate agent, real estate broker, real estate investor, real estate professional, rotarian, sales professional, small business owner \\ \hline
8 & accounting, brands, business, business development, businesses, cfp, company, consulting, cre, entrepreneurs, estate planning, experience, expert, global, helping, inc, insurance, international, investment, management, manager, managing director, managing partner, manufacturing, member finra, payroll, personal, project management, recruiting, retail, sales, services, sipc, specialist, startup, svp, talent, tax, teams, training, vice president \\ \hline
9 & account manager, brand strategist, chicagoan, communications professional, content strategist, creative thinker, data nerd, dietitian, digital marketer, digital strategist, dogmom, event planner, fundraiser, marketer, marketing consultant, marketing director, marketing manager, marketing professional, marketing specialist, marketing strategist, native new yorker, new mom, new yorker, photog, pr pro, pr professional, product designer, recruiter, social media enthusiast, social media manager, social media strategist, story teller, ui designer, urbanist, ux designer, word nerd \\ \hline
10 & act, bake, bike, camp, cook, craft, crochet, dance, draw, drink, eat, fish, game, garden, hike, hunt, knit, listen to music, lot, make, make music, paint, party, play games, play guitar, play video games, read, repeat, run, sew, shit, shop, sing, ski, sleep, stuff, surf, swim, take pictures, watch movies, work, workout, write \\ \hline
11 & acting, baking, blogging, crafting, crafts, crocheting, drawing, girls, knitting, listening to music, loves animals, loves music, painting, photography, playing video games, quilting, scrapbooking, sewing, singing, watching movies, writing \\ \hline
12 & actor, actress, arranger, artistic director, audio engineer, bass player, bassist, booking, choreographer, comedian, composer, conductor, drummer, entertainer, guitarist, improviser, itunes, lyricist, magician, model, music educator, music producer, musician, percussionist, performer, pianist, producer, promoter, rapper, recording artist, singer, singer-songwriter, skateboarder, song writer, songwriter, sound designer, violinist, vocalist \\ \hline
13 & activist, academic, anthropologist, archivist, biologist, chemist, community activist, community organizer, concerned citizen, economist, global citizen, historian, human rights advocate, humanitarian, immigrant, lawyer, lecturer, linguist, mathematician, naturalist, organizer, philosopher, physician, physicist, political activist, professor, renaissance man, researcher, scholar, scientist, social justice advocate, sociologist, translator \\ \hline
14 & adjunct professor, advocate, attorney, child advocate, college professor, communicator, community advocate, community volunteer, doctor, education advocate, mediator, ordained minister, parent, pediatrician, proud mom, public servant, retired educator, servant leader, spouse, tutor, volunteer \\ \hline
15 & active, always learning, aspiring, certified, experienced, fit, healthy, hungry, innovative, inspired, inspiring, intuitive, millennial, out, professional, small, spiritual, through, understanding \\ \hline
16 & administrator, change agent, coach, collaborator, counselor, educator, facilitator, innovator, leader, learner, mentor, motivator, presenter, teacher, visionary \\ \hline
17 & achiever, animal, animal rescuer, atheist, bernie2020, buddhist, feminist, fighter, flight attendant, gentleman, hippie, human, human being, humanist, idealist, intersectional feminist, jewish, muslim, one love, pagan, progressive, realist, rebel, socialist, truth seeker, vegan, vegetarian, warrior, witch, yanggang \\ \hline
18 & adobe, airbnb, amazon, android, apple, bitcoin, cisco, facebook, google, linkedin, microsoft, network, oracle, pinterest, product, product manager, product marketing, salesforce, twitch, twitter, uber, vmware, works \\ \hline
19 & adoption, adults, adhd, anxiety, autism, career, child, children, clinical psychologist, couples, depression, divorce, families, family therapist, marriage, mental health, mentalhealth, psychologist, psychotherapist, ptsd, relationships, students, teens, trauma, women, youth \\ \hline
20 & action, arts, birds, current events, earth, esp, history, humor, language, languages, libraries, literature, maps, museums, philosophy, plants, poetry, politics, science, science fiction, sex, space, sport, stories, trees, water, wildlife \\ \hline
21 & adventure, animals, art, books, cats, chocolate, coffee, concerts, dogs, exercise, flowers, food, gardens, good food, laughter, mountains, music, nature, people, travel, wine, yoga \\ \hline
22 & adrenaline junkie, amateur photographer, animal advocate, avid traveler, chicago native, chocolate lover, coffee snob, dog enthusiast, dog mama, dog person, dream chaser, employee, equestrian, eternal optimist, fashion lover, fashionista, fitness enthusiast, fitness fanatic, fitness junkie, fluent in sarcasm, food blogger, food enthusiast, genealogist, globetrotter, gym rat, health nut, hopeless romantic, horse lover, hr professional, jersey girl, life enthusiast, life lover, marathon runner, marketing guru, meow, minimalist, minnesotan, mom of boys, ocean lover, people lover, people person, pharmacist, pizza expert, proud mama, seeker of truth, soccer mom, social butterfly, travel addict, wanderlust, working mom, yoga enthusiast \\ \hline
23 & adventure seeker, amateur chef, anglophile, animal lover, art enthusiast, aspiring writer, avid reader, beach bum, beach lover, bibliophile, book lover, book nerd, book reader, book worm, bookworm, caffeine addict, cat lover, cat owner, chocoholic, coffee addict, coffee drinker, coffee enthusiast, coffee lover, crazy cat lady, crocheter, diyer, dog mom, food lover, foodie, future teacher, movie buff, movie lover, music enthusiast, music fanatic, music junkie, music lover, news junkie, night owl, pet lover, pop culture junkie, reality tv junkie, shopaholic, tea drinker, travel enthusiast, travel junkie, travel lover, tv junkie, voracious reader, wine drinker, wine enthusiast, wine lover, wino, workaholic \\ \hline
24 & adventures, beach, being outdoors, biking, boating, camping, cooking, cycling, dancing, eating, exploring, fishing, gardening, golfing, hiking, kayaking, laughing, loves family, motorcycles, outdoors, reading, running, sailing, shopping, skiing, snowboarding, surfing, swimming, traveling, volunteering, walking, working out \\ \hline
25 & adventurous, athletic, awkward, blonde, bold, compassionate, creative, curious, dedicated, driven, fearless, focused, free, fun loving, god fearing, good friend, goofy, grateful, hard worker, hard working, independent, motivated, nerdy, nice guy, open minded, opinionated, optimistic, passionate, positive, quirky, romantic, sarcastic, talented, tall, witty \\ \hline
26 & advisor, analyst, angel investor, board member, ceo, ceo at, cfo, chairman, cio, cmo, co-creator, co-founder, cofounder, coo, exec, executive director, founder, general manager, head, management consultant, mba, partner, past president, pmp, pres, president, treasurer \\ \hline
27 & again, atx, chi, class of 2016, cle, college, est, follow, for life, forever, htx, insta, instagram, max, now, pdx, pnw, rip, senior, snap, snapchat, team \\ \hline
28 & agent, account executive, brand ambassador, broadcaster, co-host, co-owner, columnist, coming soon, content marketer, contributor, copy editor, director, editor-in-chief, executive producer, film producer, former journalist, freelance journalist, freelancer, host, others, photojournalist, podcast, publicist, tv host, tv producer, video producer, winner \\ \hline
29 & agile, analytics, automation, aws, big data, bigdata, blockchain, cloud, compliance, crypto, cto, cybersecurity, data science, dev, devops, fintech, ibm, infosec, ios, iot, javascript, linux, machine learning, operations, product management, python, saas, security, software, solutions, startups, systems, womenintech \\ \hline
30 & agriculture, cities, climate, climate change, conservation, culture, democracy, disability, ecology, empathy, environment, environmental, equality, feminism, gender, human rights, humanity, immigration, justice, medicine, race, religion, social justice, society, transportation \\ \hline
31 & alive, amazing, ambitious, artistic, beautiful, being me, caring, classy, confident, cool, crazy, cute, determined, down to earth, easy going, educated, energetic, fabulous, friendly, fun, funny, happy, hardworking, honest, humble, in love, intelligent, just me, kind, laid back, loud, loving, loyal, nice, outgoing, outspoken, random, real, sassy, sexy, short, shy, silly, simple, smart, spontaneous, strong, sweet, unique, weird, who i am, young \\ \hline
32 & advocacy, biotech, collaboration, communication, community, csr, data, development, diversity, education, energy, engagement, entrepreneurship, ethics, finance, government, growth, healthcare, information, innovation, interested in, interests, law, leadership, neuroscience, nonprofit, nonprofits, passionate about, philanthropy, planning, policy, privacy, public policy, quality, research, safety, service, storytelling, sustainability \\ \hline
33 & all things, alabama, arizona, arkansas, alaska, atl, atlanta, austin, baltimore, bay area, born, boston, brooklyn, buffalo, cali, california, canada, charlotte, chicago, cincinnati, cleveland, colorado, columbus, d c, dallas, denver, detroit, director of sales, florida, georgia, hawaii, houston, illinois, indiana, iowa, italy, kansas, kentucky, l a, las vegas, london, long island, los angeles, louis, louisiana, madison, maine, maryland, memphis, miami, michigan, minneapolis, minnesota, nashville, native, nebraska, new jersey, new orleans, new york, new york city, nola, north carolina, nyc, office, ohio, oklahoma, oregon, orlando, paris, philadelphia, philly, phoenix, pittsburgh, portland, rva, san diego, san francisco, seattle, stl, tampa, tennessee, texas, utah, vegas, virginia, was born, washington, wisconsin, world \\ \hline
34 & all tweets, associate director, by day, gocougs, likes, links, mine, mine alone, mostly, my opinions, my own, my thoughts, my tweets, my views, not my employer's, opinions, opinions are mine, opinions mine, personal account, program manager, public affairs, retweets, rt endorsement, rts, thoughts, tweets, tweets are mine, views, views are mine \\ \hline
35 & ally, biden2020, blm, bluewave, democrat, disabled, fbr, lgbt, lgbtq, liberal, mark twain, metoo, nevertrump, no dms, notmypresident, proud democrat, resist, resistance, resister, theresistance, uniteblue, vote, voteblue, votebluenomatterwho \\ \hline
36 & alum, board, columbia, comms, cornell, current, currently, dukeu, fellow, fmr, for, former, formerly, georgetown, harvard, intern, member, mit, northwesternu, past, penn, prev, previously, princeton, stanford, ucberkeley, uchicago, umich, uwmadison, via, yahoo \\ \hline
37 & alumna, alumni, alumnus, anchor, bgsu, cal, ella, grad, graduate, journalism, mizzou, mpa, ohiostate, ohiou, proud, resident, rts endorsements, sdsu, syracuseu, ucla, uconn, unc, uofsc, usc, utaustin, uva \\ \hline
38 & always, also, all, ask me, bad, best, but, cheers, don't worry, dude, end, ever, every day, everything, good, guy, hello, here, hey, in the end, it's me, know, like, livin' the dream, make things, man, maya angelou, new to twitter, no regrets, not, okay, one, only, please, really, right, see, seriously, sometimes, thank you, thanks, that, that's me, then, there, things, this is me, too, visit, wait, well, what, will, yeah, yep, yes, you, you know, yup \\ \hline
39 & accessories, antiques, color, diy, everything in between, fashion, films, hair, hugs, interior design, jewelry, lover of art, makeup, modeling, much more, pets, quotes, skincare, smiles, spa, spirits, vintage, welcome \\ \hline
40 & all around nerd, avid gamer, cat dad, cinephile, cosplayer, dork, gamer, geek, martial artist, metal head, metalhead, movie, nerd, pc gamer, streamer, trekkie, twitch affiliate, twitch streamer, video game enthusiast, video gamer, whovian, wrestling fan \\ \hline
41 & ambassador, 5th grade teacher, apple teacher, assistant principal, doctoral student, elementary principal, google certified educator, graduate student, history teacher, instruction, instructional coach, instructional designer, life long learner, life-long learner, lifelong learner, lover of learning, m ed, math teacher, nbct, passionate educator, principal, science teacher, spanish teacher, superintendent \\ \hline
42 & amen, enjoying life, family is everything, god bless, god is good, god is love, jesus christ, just being me, let live, life is good, life is great, livin, living life, lord, lover of god, lover of jesus, loves god, loves jesus, loves life, lovin life, lovin' life, loving god, loving life, loving my life, madness is genius, my job, my life, myself, savior, small town girl, working hard, worship \\ \hline
43 & america, army, blue, country, election2016, fire, israel, liberty, military, navy, ret, u s, us army, usa, usaf, usmc, veterans \\ \hline
44 & among other things, amwriting, aspiring author, albert einstein, book, creative writer, fiction, fiction writer, freelance editor, link below, lover of words, out now, rep, rep'd by, repped by, reviewer, romance, scbwi, technical writer, writingcommunity \\ \hline
45 & 49ers, angels, arsenal, astros, badgers, bears, bengals, bills, blackhawks, braves, brewers, broncos, browns, bruins, buckeyes, bucks, bucs, bulls, caps, cardinals, cavs, celtics, chiefs, colts, cowboys, cubs, dallascowboys, dodgers, dolphins, ducks, duke, eagles, falcons, fan, flyers, gators, giants, hawkeyes, hawks, heat, huge sports fan, huskers, indians, kings, knicks, lakers, lfc, mariners, mets, notre dame, nyy, orioles, pacers, packers, panthers, penguins, penn state, pens, phillies, pirates, raiders, rangers, ravens, red sox, reds, redskins, redsox, royals, sabres, saints, seahawks, sharks, sixers, spurs, steelers, texans, tigers, titans, twins, usmnt, vikings, vols, warriors, yankees \\ \hline
46 & american, army vet, army veteran, businessman, catholic, christian, citizen, college grad, constitutional conservative, constitutionalist, free thinker, grandparent, independent thinker, libertarian, navy veteran, pro-life, proud american, republican, texan, vet, veteran \\ \hline
47 & animator, art director, artist, bon vivant, cartoonist, cinematographer, comic, content creator, costume designer, creative director, creator, designer, digital artist, film maker, filmmaker, fine artist, freelance graphic designer, game designer, graphic artist, graphic designer, illustrator, music maker, painter, photographer, podcaster, printmaker, raconteur, sculptor, video editor, videographer, visual artist, voice actor, web designer \\ \hline
48 & animation, advertising, audio, computer, editing, entertainment, film, freelance, graphic, graphic design, graphics, illustration, news, photo, photos, pop culture, print, production, promotions, publishing, radio, television, video, video production, videos, web design, websites \\ \hline
49 & anime, batman, beatles, cosplay, d\&d, doctor who, fantasy, harry potter, horror, lost, manga, marvel, outlander, paranormal, sci-fi, scifi, star trek, star wars, starwars, supernatural, twd, walking dead \\ \hline
50 & aggie, baseball coach, basketball coach, civil engineer, devoted husband, eagle scout, father of 2, father of 3, father of 4, father of 5, father of four, father of three, football coach, happy husband, husband of 1, husband to, loving father, loving husband, lucky husband, married to, nascar fan, navy vet, player, proud dad, proud father, proud husband, soccer coach, social studies teacher, soldier, steelers fan \\ \hline
51 & aquarius, aries, bisexual, bitch, black, cancer, capricorn, college student, female, free spirit, future mrs, gay, gemini, italian, leo, lesbian, libra, male, mexican, pisces, positive vibes, princess, puerto rican, queen, sagittarius, scorpio, tattooed, taurus, tired, virgo, white \\ \hline
52 & architect, audiophile, coder, computer scientist, data scientist, developer, engineer, game developer, guitar player, hacker, maker, motorcyclist, programmer, software developer, software engineer, tech enthusiast, tech nerd, techie, technologist, technology enthusiast, technophile, tinkerer, web developer, woodworker \\ \hline
53 & architecture, building, code, coding, design, electronics, internet, investing, mobile, networking, programming, sound, tech, technology, web, web development \\ \hline
54 & adventurer, art lover, baker, dog lover, eater, environmentalist, explorer, francophile, gardener, knitter, librarian, nature lover, optimist, quilter, reader, runner, scrapbooker, student of life, traveler, traveller, tree hugger, walker, wanderer, world traveler, yogi, yogini \\ \hline
55 & army wife, best friend, blessed beyond measure, breast cancer survivor, business woman, busy, cancer survivor, caregiver, christian wife, devoted wife, fields, full time mom, full time student, grammy, granddaughter, grandma, grandmom, homemaker, housewife, lover of people, loving mother, loving wife, loyal friend, mom of 4, mom of 5, mommy of 2, most importantly, mother of 3, mother of 4, mother of 5, mother of four, mother of three, mother of two, nurse practitioner, paralegal, proud mother, registered nurse, retired nurse, retired teacher, rodan, sahm, single mother, so much more, survivor, widow, wife of 1, woman, woman of god, work in progress \\ \hline
56 & art teacher, barista, bartender, colorist, disney fanatic, esthetician, fashion designer, freelance photographer, hair stylist, hairstylist, interior designer, jewelry designer, jewelry maker, lifestyle blogger, makeup artist, mua, on instagram, portrait photographer, professional photographer, salon owner, seamstress, stylist, tattoo artist, vlogger, wedding photographer, youtuber \\ \hline
57 & artists, biz, blog, by night, check out, com, contact, email, etc, etsy, gmail, inquiries, medium, more, new, spotify, stay tuned, tips, voice, website, with, words, writer for, youtube \\ \hline
58 & associates, abr, auto, broker, buy, buyers, buying, call, commercial, commercial real estate, construction, cosmetic, farm, gri, home, investments, land, llc, local, luxury, nmls, own, owner, real estate, realestate, realtor, realtors, residential, sell, sellers, selling, surrounding areas \\ \hline
59 & astronomy, board games, broadway, cars, cartoons, comedy, comic books, comics, computers, disney, fast cars, gadgets, games, gaming, genealogy, guitars, horror movies, jokes, martial arts, movies, musicals, podcasts, skateboarding, sports, tattoos, theater, theatre, tv shows, video games, videogames, woodworking, zombies \\ \hline
60 & atc, athletic trainer, bbn, beardown, boomer sooner, bred, counting, cowboysnation, cscs, family first, field, geaux tigers, go cougs, go dawgs, go gators, go hawks, go pack go, go vols, godawgs, gohawks, gopackgo, jeremiah 29:11, john 3:16, joshua 1:9, jucoproduct, kcco, ksu, lover of sports, mffl, mia, osu, pe teacher, phil 4:13, philippians 4:13, proverbs 3:5-6, r i p, raised, roll tide, romans 8:28, seuss, texas a\&m, texas born, texas forever, track, ttu, uga, university of alabama, vfl, war eagle, who dat, wps, wsu, zta \\ \hline
61 & athlete, boss, cat enthusiast, cheerleader, collector, crafter, dancer, dog-lover, dreamer, drinker, goofball, listener, lover, mixed media artist, mover, ninja, opera singer, person, planner, potter, procrastinator, student, superhero, thinker, weirdo, wordsmith, worker \\ \hline
62 & athletics, avid, avid sports fan, billsmafia, byu, chiefskingdom, coys, fanatic, flyeaglesfly, fsu, fsu alum, gbr, girldad, go blue, go bucks, goblue, goducks, gold, h2p, hailstate, head coach, httr, illini, jets, keeppounding, lsu, mlb, nyr, proud member, psu, raidernation, rolltide, sfgiants, skol, sneakerhead, supporter, unlv, whodat, wvu, ynwa \\ \hline
63 & auntie, aunt, child of god, cousin, daughter, friend, girlfriend, grandmother, homeschooler, jesus lover, mama, mimi, mom, momma, mommy, mother, nana, niece, nurse, pastor's wife, sister, stepmom, wife, wifey \\ \hline
64 & author, best selling author, business coach, business consultant, career coach, consultant, evangelist, executive coach, influencer, inspirational speaker, instructor, keynote speaker, leadership coach, life coach, motivational speaker, practitioner, professional speaker, public speaker, published author, radio host, seeker, speaker, trainer \\ \hline
65 & acupuncturist, beachbody coach, certified personal trainer, doula, fitness coach, fitness instructor, healer, health coach, herbalist, join me, licensed massage therapist, lover of life, lover of nature, massage therapist, meditator, mompreneur, mother of twins, namaste, nutritionist, occupational therapist, personal trainer, registered dietitian, reiki master, therapist, veterinarian, wellness coach, yoga instructor, yoga teacher \\ \hline
66 & acab, black lives matter, blacklivesmatter, her, hers, him, his, pronouns: she, queer, she, them, they \\ \hline
67 & aviation, bass, cashapp, drums, edm, guitar, hip hop, hip-hop, jazz, karaoke, mac, memes, metal, netflix, of course, opera, other stuff, other things, piano, pop, ps4, r\&b, rap, rock, roll, scuba, sup, weather, whatever, wow, xbox \\ \hline
68 & avid cyclist, banker, bowler, car enthusiast, computer geek, computer programmer, data analyst, electrical engineer, family guy, history buff, hockey fan, hockey player, home brewer, homebrewer, ironman, it professional, master of none, mechanical engineer, motorcycle enthusiast, poker player, private pilot, science nerd, skeptic, soccer fan, soccer player, tech geek, tech junkie, urban planner \\ \hline
69 & avid golfer, baseball fan, beer enthusiast, beer snob, boston sports fan, buckeye, chicago sports fan, cleveland sports fan, colorado native, craft beer enthusiast, cubs fan, dog dad, football fan, football fanatic, history nerd, hoosier, husker fan, law student, meteorologist, mets fan, michigander, music fan, native texan, outdoors enthusiast, packer fan, physical therapist, political junkie, politics junkie, pop culture enthusiast, proud parent, red sox fan, saints fan, spartan, sports enthusiast, sports fan, sports fanatic, sports junkie, sports lover, sports nut, tar heel, yankees fan \\ \hline
70 & avid runner, backpacker, beekeeper, beer drinker, beer lover, bicyclist, biker, birder, boater, brewer, camper, chef, climber, conservationist, crossfitter, cyclist, dog owner, geologist, hiker, kayaker, marathoner, mountain biker, outdoor enthusiast, rock climber, sailor, scuba diver, skier, snowboarder, surfer, swimmer, tennis player, trail runner, triathlete \\ \hline
71 & b2b, brand, branding, communications, consumer, content, content marketing, contentmarketing, digital, digital marketing, digital media, digitalmarketing, ecommerce, fundraising, marketing, media, media relations, online, partnerships, ppc, public relations, seo, social, social media, social media marketing, socialmedia, strategy, wordpress \\ \hline
72 & babies, beyond, corporate, editorial, event, event planning, events, fine art, hospitality, landscape, lifestyle, oil, portrait, portraits, special events, wedding, weddings, workshops \\ \hline
73 & backtheblue, conservative, constitution, god bless america, kag, kag2020, maga, my country, nra, patriot, prolife, qanon, trump, trump 2020, trump supporter, trump2020, wwg1wga \\ \hline
74 & bacon, beaches, bikes, boston sports, bourbon, cheese, cocktails, cookies, craft beer, cupcakes, glitter, good beer, good book, good books, good friends, good music, good people, good times, good vibes, good wine, great food, great friends, great outdoors, ice cream, live music, lover of animals, lover of books, lover of cats, lover of coffee, lover of dogs, lover of family, lover of food, lover of music, lover of travel, lover of wine, loves, most of all, my cat, naps, ocean, pizza, puns, puppies, reality tv, red wine, sarcasm, scotch, sea, shoes, snow, starbucks, summer, sun, sunsets, sunshine, sushi, tacos, target, tea, turtles, whiskey, zumba \\ \hline
75 & barber, blessed, bsn, carpe diem, cna, cosmetologist, crafty, dog, emt, engaged, er nurse, for now, forgiven, future nurse, girl, god 1st, god first, hairdresser, hakuna matata, highly favored, jesus freak, loved, medical assistant, nanny, nursing student, proud mommy, proverbs 31:25, psalm 46:5, redeemed, redhead, rocket ship builder, saved, saved by grace, single mom, sinner, southern, teamjesus, thankful, truly blessed, twin \\ \hline
76 & baseball, basketball, bbq, beer, bowling, boxing, cigars, classic rock, college football, crossfit, football, golf, green bay packers, gym, hockey, hunting, lacrosse, mma, poker, racing, rugby, soccer, softball, tennis, trucks, volleyball, wrestling \\ \hline
77 & be happy, be kind, be strong, be you, be yourself, believe, believe in yourself, breathe, build, create, do good, dream, dream big, enjoy, enjoy life, explore, grow, have courage, have fun, inspire, keep moving forward, laugh, laugh a lot, laugh often, learn, listen, live, live it, live to love, live well, meet new people, never give up, play, play hard, play harder, pray, relax, share, smile, stay humble, stay positive, talk, teach, think, tweet, work hard \\ \hline
78 & beauty, body, coaching, creativity, fitness, healing, health, healthy living, ideas, inspiration, meditation, mindfulness, motherhood, movement, nutrition, organic, parenting, pilates, places, products, recipes, reiki, spirit, spirituality, style, wellness \\ \hline
79 & believer, brother, chaplain, christ follower, christ-follower, church planter, disciple, disciple of christ, encourager, follower, follower of christ, follower of jesus, friend to many, husband of one, jesus follower, man of god, minister, missionary, music teacher, pastor, preacher, senior pastor, servant, son, uncle, worship leader, worshiper, worshipper, youth pastor \\ \hline
80 & bestselling author, blogger, children's book author, copywriter, critic, curator, editor, essayist, freelance writer, humorist, novelist, observer, playwright, podcast host, poet, publisher, radio personality, screenwriter, storyteller, travel writer, writer \\ \hline
81 & better, creating, everyday, farming, growing, having fun, helping others, helping people, learning, learning new things, lewis, listening, living, meeting new people, playing, school, sharing, teaching, traveling the world, working \\ \hline
82 & biology, chemistry, computer science, economics, edtech, engineering, french, gis, literacy, math, performance, physics, psychology, robotics, stem, theology \\ \hline
83 & blues, chicago cubs, dallas cowboys, denver broncos, lions, louis cardinals, magic, nascar, nba, nfl, nhl, nuff said, ny giants, patriots, pro wrestling, sf giants, stars, ufc, wwe \\ \hline
84 & builder, business analyst, community builder, connector, doer, entrepreneur, executive, futurist, inventor, investor, networker, philanthropist, problem solver, project manager, serial entrepreneur, social entrepreneur, strategist, thought leader, trader \\ \hline
85 & care, change, compassion, connect, courage, dignity, dreams, educate, empower, faith, freedom, goals, grace, gratitude, happiness, hard work, heart, help, honesty, hope, integrity, joy, kindness, knowledge, life, light, love, loyalty, mind, money, motivation, passion, passionate about life, peace, positivity, power, purpose, respect, soul, strength, success, support, trust, truth, unknown, wealth, wisdom \\ \hline
86 & carpenter, chiropractor, driver, electrician, farmer, firefighter, it guy, marine, mechanic, mormon, papa, paramedic, salesman, truck driver, us army veteran, usaf veteran, volunteer firefighter, welder \\ \hline
87 & cat lady, cat mom, cat person, daydreamer, empath, enfj, enfp, fangirl, ginger, grad student, gryffindor, hufflepuff, infj, infp, intj, introvert, mama bear, mental health advocate, old soul, pastry chef, ravenclaw, slytherin, social worker, spoonie \\ \hline
88 & christ, church, country music, family, friends, god, grandchildren, grandkids, guns, horses, jesus, kids, my children, my daughter, my dog, my dogs, my family, my friends, my god, my hubby, my husband, my kids, my son, my wife \\ \hline
89 & dad, daddy, family man, father, father of two, fisherman, golfer, grandfather, grandpa, hunter, husband, outdoorsman, pilot \\ \hline
    \caption{Caption}
    \label{tab:app_clust}
\end{longtable}

\section{Meaning Differences for Pre-Defined Lists of Terms for Panel 2 and Random}

	\begin{figure}[h]
		\centering
	     \begin{tabular}{c c}
		      \includegraphics[width=.5\textwidth]{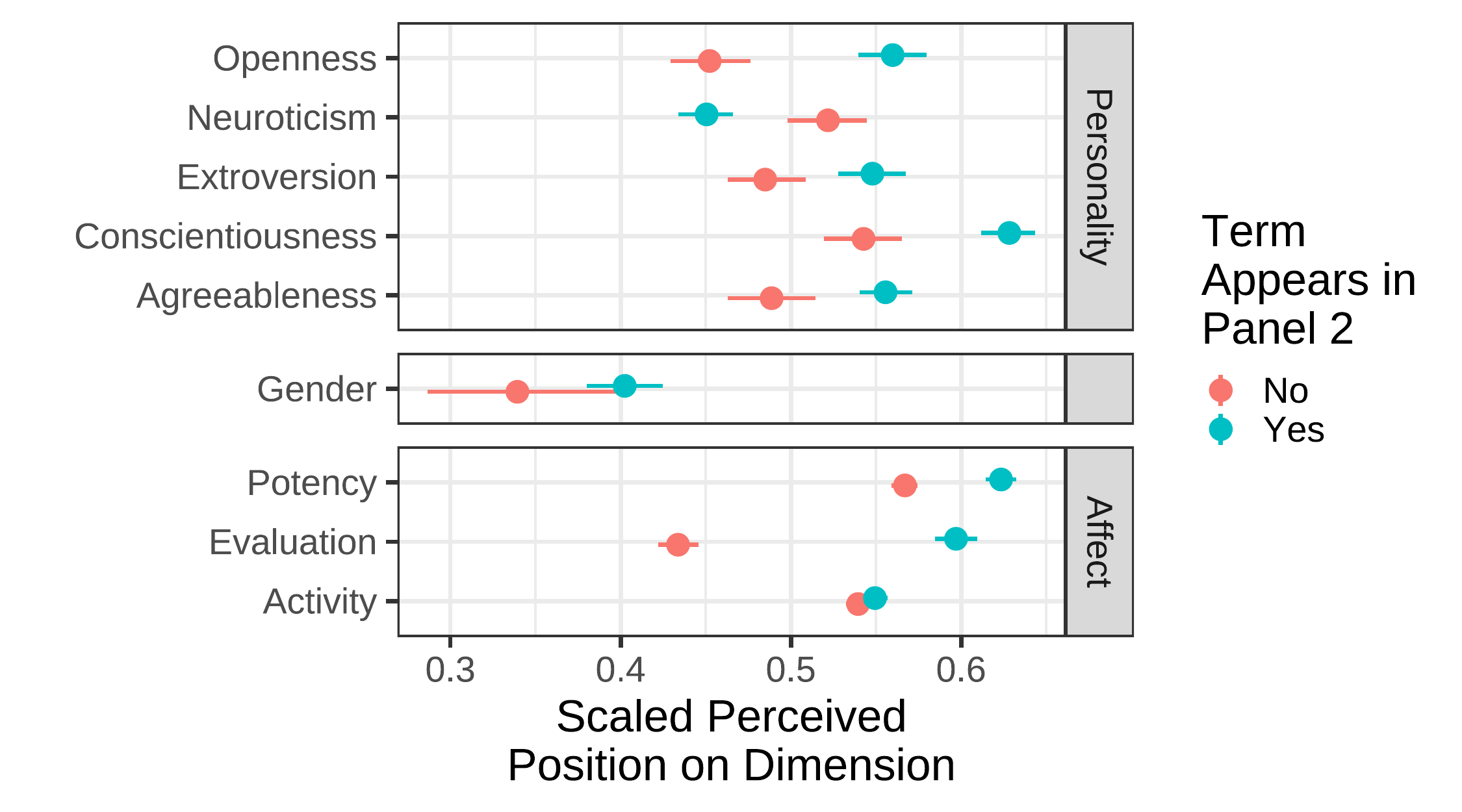}   & 
		      \includegraphics[width=.5\textwidth]{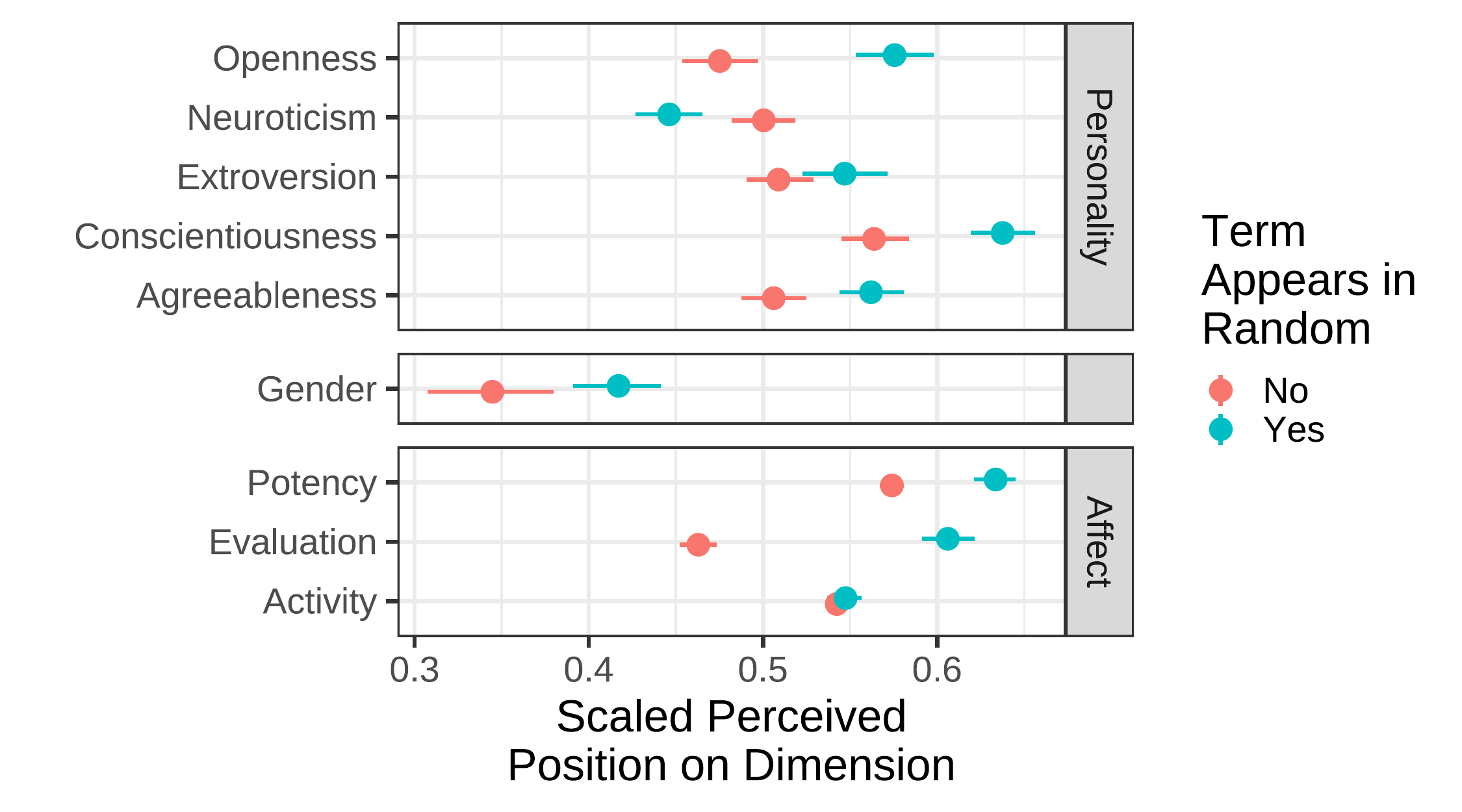} \\
		         (a) & (b)
		    \end{tabular}
		
		\caption{Results for Panel 2 (a) and Random (b) for the analysis presented in Figure~\ref{fig:lists}b)}
		\label{fig:lists_appendix}
	\end{figure}

\end{document}